\pdfoutput=1

\documentclass[fleqn,usenatbib]{mnras}

\usepackage{mathptmx}
\usepackage[T1]{fontenc}
\usepackage{ae,aecompl}

\usepackage{graphicx}	% Including figure files
\usepackage{subfigure}
\usepackage{amssymb}	% Extra maths symbols
\usepackage{amsmath}	% Advanced maths commands

\newcommand{\angstrom}{\text{\normalfont\AA}}

\title[The assembly history of NGC 3115]{The assembly history of the nearest S0 galaxy NGC 3115 from its kinematics out to six half-light radii}

\author[Dolfi et al.]{Arianna Dolfi,$^{1}$\thanks{E-mail: adolfi@swin.edu.au}
Duncan A. Forbes,$^{1}$
Warrick J. Couch,$^{1}$
Anna Ferr\'e-Mateu,$^{2,1}$
\and Sabine Bellstedt,$^{3}$
Kenji Bekki,$^{3}$
Jonathan Diaz,$^{3}$
Aaron J. Romanowsky,$^{4,5}$
\and Jean P. Brodie,$^{5}$
\\
\\
$^{1}$ Centre for Astrophysics \& Supercomputing, Swinburne University of Technology, Hawthorn VIC 3122, Australia\\
$^{2}$ Institut de Ciencies del Cosmos (ICCUB), Universitat de Barcelona (IEEC-UB), E02028 Barcelona, Spain\\
$^{3}$ ICRAR, M468, The University of Western Australia 35 Stirling Highway, Crawley Western Australia, 6009, Australia\\
$^{4}$ Department of Physics $\&$ Astronomy, San Jos\'e State University, One Washington Square, San Jose, CA 95192, USA\\
$^{5}$ University of California Observatories, 1156 High St., Santa Cruz, CA 95064, USA\\
}

\date{Accepted 2020 April 9. Received 2020 April 9; in original form 2020 February 26}

\pubyear{2020}

\begin{document}
\label{firstpage}
\pagerange{\pageref{firstpage}--\pageref{lastpage}}
\maketitle

\begin{abstract}
Using new and archival data, we study the kinematic properties of the nearest field S0 galaxy, NGC 3115, out to $\sim6.5$ half-light radii ($R_\mathrm{e}$) from its stars (integrated starlight), globular clusters (GCs) and planetary nebulae (PNe). We find evidence of three kinematic regions with an inner transition at $\sim0.2\ R_\mathrm{e}$ from a dispersion-dominated bulge ($V_\mathrm{rot}/\sigma <1$) to a fast-rotating disk ($V_\mathrm{rot}/\sigma >1$), and then an additional transition from the disk to a slowly rotating spheroid at $\sim2-2.5\, R_\mathrm{e}$, as traced by the red GCs and PNe (and possibly by the blue GCs beyond $\sim5\, R_\mathrm{e}$). From comparison with simulations, we propose an assembly history in which the original progenitor spiral galaxy undergoes a gas-rich minor merger that results in the embedded kinematically cold disk that we see today in NGC 3115. At a later stage, dwarf galaxies, in mini mergers (mass-ratio $<$ 1:10), were accreted building-up the outer slowly rotating spheroid, with the central disk kinematics largely unaltered. Additionally, we report new spectroscopic observations of a sample of ultra-compact dwarfs (UCDs) around NGC 3115 with the Keck/KCWI instrument. We find that five UCDs are inconsistent with the general rotation field of the GCs, suggesting an \textit{ex-situ} origin for these objects, i.e. perhaps the remnants of tidally stripped dwarfs. A further seven UCDs follow the GC rotation pattern, suggesting an \textit{in-situ} origin and, possibly a GC-like nature.
\end{abstract}


%%%%%%%%%%%%%%%%%%%%%%%%%%%%%%%%%%%%%%%%%%%%%%%%%%

%%%%%%%%%%%%%%%%% BODY OF PAPER %%%%%%%%%%%%%%%%%%

% --------------- Introduction Section ---------------- %
\section{Introduction}
The formation of lenticular (S0) galaxies still remains an open question in astrophysics. The S0 morphological classification was first introduced by \citet{Hubble1926}, who placed them as the connecting link between spirals and ellipticals in his tuning fork diagram because they share similarities with both spiral and elliptical galaxies, being composed of a central bulge component and a surrounding disk that lacks spiral arms. 

S0 galaxies are observed in much larger numbers in the local Universe than at high-redshifts, as well as towards the centre of clusters at the expense of spirals \citep{Dressler1980,Dressler1997}. This evidence, combined with their morphological structure, has long suggested that S0s could be an evolution of spiral galaxies that quenched their star-formation and lost their spiral arms due to the interaction with the dense environment (e.g. \citealt{Butcher1978a,Butcher1978b}).
The physical processes responsible for this may involve ram-pressure stripping, tidal interactions or galaxy harassment (e.g. \citealt{Gunn1972,Larson1980,Bekki2009,Bekki2011,Merluzzi2016}). All these processes can explain the presence of S0s in cluster environments, but not in the field, where interactions with the environment are limited, if non-existent.
Thus, the presence of field S0s widens the range of possible formation scenarios, such as past or recent major/minor mergers, gas accretion onto compact ellipticals or secular evolution processes, such as disc instabilities or
feedback events (e.g. \citealt{Bekki1998,Bournaud2005,Naab2014,Bassett2017,Bellstedt2017,Saha2018,Diaz2018,Rizzo2018,Mishra2018,ElicheMoral2018}). 

One commonly proposed scenario for the formation of massive early-type galaxies (ETGs) 
is the two-phase model, e.g. \citet{Oser2010,Rodriguez2016}. According to the model, these galaxies form in an early phase of rapid collapse or major merger at $z\geq2$, during which they form most of their stellar mass from the in-falling cold gas towards the central regions (\textit{in-situ} component), and at later epochs, $z\leq2$, they keep growing mass from the accretion of satellite galaxies onto the outer regions via minor mergers (\textit{ex-situ} component). Hence, it is expected to find stellar kinematic and stellar population transitions within a galaxy at different radii (e.g. \citealt{Wu2014,Cook2016,AFM2019}). 

Recent integral field spectroscopy (IFS) surveys, e.g. SAURON \citep{SAURON2001}, ATLAS$^{\rm 3D}$ \citep{ATLAS3D2011}, SAMI \citep{SAMI2012}, MASSIVE \citep{MASSIVE2014}, MaNGA \citep{MaNGA2015} and CALIFA \citep{CALIFA2016}, as well as multi-object spectrograph surveys, e.g. SLUGGS \citep{Brodie2014}, have been revolutionary in obtaining stellar kinematics, as well as spatially resolved stellar populations, of large samples of early-type galaxies.  
From the stellar population analysis of the MaNGA S0 galaxy sample, \citet{Fraser2018} found that the main discriminator in the formation of S0s is their mass, instead of the environment in which they are located. They suggested that massive S0s, with an older and more metal-rich stellar population in the bulge than in the disk, formed through an inside-out quenching scenario, while low-mass S0s, with younger bulges than the disks, formed through an outside-in quenching scenario (faded spirals), involving episodes of gas accretion and inflow that induced star-formation events near the bulge. 

On the contrary, \citet{Coccato2020} found that cluster S0s are more rotationally supported (consistent with a faded spiral scenario), while field S0s are more pressure supported (minor mergers influence) from their spatially resolved kinematic studies of a sample of $21$ cluster and field S0s. This suggests that there is a wide range of physical processes playing a role in the formation of S0 galaxies, which depend on environment. However, they did not find the same dependence on the environment from the stellar population properties of their S0 sample.

From a stellar kinematic analysis, as part of the SLUGGS survey, \citet{Bellstedt2017} suggested that low-mass S0s are likely formed through secular processes from fading spirals rather than mergers, although recent merger events could not be excluded. Based on early SLUGGS data, \citet{Arnold2014} found evidence of fast-rotating disks embedded in slower rotating spheroidal components, suggesting a two-phase model for most of their early-type galaxies. \citet{Foster2016} also found evidence of embedded disks in their observed stellar kinematics of early-type galaxies in the SLUGGS survey, consistent with a two-phase model. \citet{Foster2018} found only a very small fraction of embedded disk structures from the stellar kinematics of galaxies from the SAMI survey. However, they suggested that this could be due to sample selection effects or to the limited extension of their kinematic data not reaching beyond $\sim2.5\, R_\mathrm{e}$.
\citet{Forbes2016} found that most of the SLUGGS early-type galaxies were formed through gas-rich minor mergers and were characterized by a mass growth of the outer regions at later epochs, which supports a two-phase model.

Stellar kinematic studies of ETGs are limited as they usually do not extend beyond $\sim1-2\, R_\mathrm{e}$ or $\sim3-4\, R_\mathrm{e}$ in the case of the SLUGGS survey. This is because the stellar surface brightness decreases very steeply and, as a result, it becomes difficult and time-consuming to observe at large radii.
The galaxy outskirts enclose large fractions of the galaxy stellar mass, angular momentum and accretion signatures (see e.g. figure 1 in \citealt{Brodie2014}) and are, thus, highly relevant for galaxy formation and evolution studies (e.g. angular momentum as discussed in \citealt{Romanowsky2012}).
For this reason, discrete tracers such as globular clusters (GCs) and planetary nebulae (PNe) have been used as probes of the outskirts of galaxies as they are more luminous and, thus, can be observed out to larger radii, i.e. $\sim8-10\, R_\mathrm{e}$, than the starlight.
Recent surveys were able to obtain photometric and spectroscopic measurements for a large number of GC and PNe systems around early-type galaxies, namely SLUGGS \citep{Brodie2014,Forbes2017} and the extended Planetary Nebula Spectrograph Survey (ePN.S, \citealt{Pulsoni2018Survey}). 

Many works have used these large GC and PNe systems to investigate the kinematic properties of early-type galaxies out to large galactocentric radii as they are both found to be good tracers of the underlying stellar population and stellar surface brightness profile (e.g. \citealt{Romanowsky2006,Coccato2009,McNeil2010,Arnold2011,Cortesi2011,Cortesi2013b,Coccato2013,Pota2013,Cortesi2016,Hartke2017,Zanatta2018,Forbes2018}).  
Simulations by \citet{Bekki2005} found that major mergers can significantly boost the rotation of both the red and blue GC sub-populations out to large radii, while minor mergers are responsible for increasing the random motion in the outer regions of galaxies, causing the corresponding fall-off of the $V_\mathrm{rot}/\sigma$ profile which is observed in some S0 galaxies at large radii (e.g. \citealt{Arnold2011,Zanatta2018}).

Another class of compact stellar objects, which can be useful kinematic tracers of galaxies out to large radii, are the ultra-compact dwarfs (UCDs).
UCDs were first discovered in the centre of the Fornax cluster \citep{Hilker1999,Drinkwater2000}, and were later observed in a range of environments \citep{Chillingarian2010,Jennings2014,Penny2014,Liu2015,Zhang2015,Voggel2016,Forbes2017}.
UCDs are defined to be more massive, more luminous and larger than typical GCs. From previous studies, they are characterized by dynamical masses $10^{6}\, \mathrm{M_{\odot}}<M^{*}<10^{9}\, \mathrm{M_{\odot}}$ (i.e. $\sigma\sim20-50\, \mathrm{km\,s^{-1}}$), half-light radius $10\, \mathrm{pc} < R_{h} < 100\, \mathrm{pc}$ and typical absolute V-magnitudes $M_{V} < -10.25\, \mathrm{mag}$ \citep{Hilker2007a,Hilker2007b,Chillingarian2010,Mieske2012,Norris2014}.
Observations have found a range of UCDs with slightly different properties between each other, thus suggesting that some are consistent with being massive star-clusters, while others are more consistent with nucleated dwarf galaxies which have been tidally stripped (e.g. \citealt{Bekki2003,Hilker2007a,Hilker2007b,Frank2011,Liu2015}). 
Simulations have found that UCDs may be the product of tidally stripped nucleated dwarf galaxies (e.g. \citealt{Bekki2003,Bekki2015,Pfeffer2016,Goodman2018}) 
and that the range of physical sizes observed for UCDs is a result of the different possible orbits and  number of pericentric passages of the dwarf galaxy \citep{Pfeffer2013}.

In this work, we perform new Keck/KCWI spectroscopic observations of a sample of UCDs around the nearest field S0 galaxy, NGC 3115, which were initially identified with {\it HST} photometry by \citet{Jennings2014} (hereafter J14). We aim to understand the origin of the UCDs in NGC 3115 from the comparison with the velocity field of the GCs. For this purpose, we study the kinematic properties of NGC 3115 by combining the stellar kinematics from SLUGGS \citep{Arnold2011,Arnold2014} and VLT/MUSE \citep{Guerou2016} with GCs \citep{Arnold2011,Pota2013,Jennings2014,Forbes2017} and, recently observed, PNe \citep{Pulsoni2018Survey} systems. In \citet{Zanatta2018}, they found evidence of a disk-like to spheroid-like transition with radius in NGC 3115 by studying the kinematics of its GC and PNe systems and they proposed a series of minor-mergers as the likely formation scenario for this galaxy.
The large galactocentric radii  ( i.e. $\sim6.5\, R_\mathrm{e}$) probed by combining the above kinematic datasets  allows us to study the presence of distinct kinematic transitions within NGC 3115. From this we describe a possible formation pathway for NGC 3115 and the origin of its UCDs. 

This paper is structured as follows: in Sec. \ref{sec:sample_selection} to \ref{sec:data_reduction}, we describe our selected sample of UCD candidates from J14, our observations, and the data reduction. In Sec. \ref{sec:data_analysis}, we describe the kinematic analysis of our observed UCDs. In Sec. \ref{sec:TheGalaxy}, we describe our kinematic datasets for NGC 3115, namely the SLUGGS integrated starlight (hereafter SLUGGS stars), GCs and PNe.
In Sec. \ref{sec:PointSymmetry}, we test the symmetry of our kinematic datasets. 
In Sec. \ref{sec:results}, we present our kinematic results. Here, we also compare our measured UCD velocities with the underlying GC velocity field. In Sec. \ref{sec:Discussions}, we discuss our kinematic results and propose a formation scenario for NGC 3115 based on comparison with simulations of galaxy formation. In Appendix \ref{sec:UCD_candidates_images}, we show the {\it HST} images of the $31$ UCDs listed in J14. In Appendix \ref{sec:background_objects}, we show the UCDs identified as background galaxies from our spectroscopic analysis and, finally, in Appendix \ref{sec:issues}, we briefly summarize some issues we found with the UCDs described in J14.
Throughout the paper we adopt $d=9.4\, \mathrm{Mpc}$ as the distance to NGC 3115 \citep{Brodie2014}. 

% ------------------ Sample Selection Section -------------------- %
\section{Data}
\label{sec:Data}

\subsection{UCD Sample Selection}
\label{sec:sample_selection}
In this work we select our sample of NGC 3115 UCD candidates from J14. 
J14 studied the GC system of NGC 3115 using the {\it Hubble Space Telescope}/Advanced Camera for Surveys ({\it HST}/ACS). They obtained photometric $g$- and $z$- magnitudes and half-light radii measurements for $360$ GCs. 
Among these objects, they also identified $31$ UCD candidates around NGC 3115 (see J14, table 3). 
To distinguish between the UCD candidates and the GCs, size was used as the main selection criterion, i.e. half-light radius $R_{h} > 8\, \mathrm{pc}$, for the UCDs under the assumption that they lie at the NGC 3115 distance. 
We note, however, that \citet{Cantiello2015} pointed out some issues with the J14 UCD selection criteria, such as the presence of clear background contaminants in their sample of $31$ selected UCDs. Therefore, we refine the UCD selection process, initially performed by \citet{Jennings2014}, by visually inspecting the {\it HST}/ACS mosaic images of the field around NGC 3115, where all $31$ UCD candidates lie. In this way, we select a final sample of UCD candidates for observation and we reject all other objects as background contaminants based on the following visual selection criteria:

\begin{itemize}
    \item they have a shape significantly different from a round spherical GC-like one;
    \item they have clear features that suggest a background galaxy, e.g. spiral arms;
    \item they lie very close to the main galaxy, NGC 3115, and they are thus heavily contaminated by the galaxy light.
\end{itemize}

We exclude the objects satisfying the latter criterion because they would be hard to observe and isolate from the background galaxy light.
Figures \ref{fig:GoodUCDs} and \ref{fig:BadUCDs} of Appendix \ref{sec:UCD_candidates_images} show snapshots of the $13$ selected and $13$ rejected UCD candidates, respectively. The remaining $5$ UCDs are shown separately in Fig. \ref{fig:GoodUCDsVel} since these objects were already spectroscopically confirmed members of NGC 3115 from their radial velocity measurements \citep{Pota2013}. 
In Table \ref{tab:UCD_Candidates_Table}, we show the list of all $31$ UCD candidates.

\begin{table}
\centering
\caption{List of all $31$ UCDs from J14. \textit{Column 1:} UCD ID as in J14. \textit{Column 2, 3:} new RA and DEC coordinates centered on each UCD from the {\it HST}/ACS $g$-band images. \textit{Column 4:} apparent $g$-magnitude of the UCD taken from J14. \textit{Column 5:} total exposure times of the UCD that we observed with KCWI (see Sec. \ref{sec:Observations}). UCD 17a is the ``bonus" object that we obtained during our KCWI observations (Sec. \ref{sec:Observations}), which lies close to UCD 17 (see Fig. \ref{fig:GoodUCDs}). We estimate its $g$ magnitude from the {\it HST}/ACS images using the \textit{qphot} task within IRAF.}
\begin{tabular}{|c|c|c|c|c|}\hline\hline
UCD & $\mathrm{RA}$ ($\degr$) & $\mathrm{DEC}$ ($\degr$) & $m_{g}$ (mag) & Exp. Time ($\mathrm{s}$) \\ \hline
1 & 151.3021600 & $-$7.7355695   & $19.118$ & $360$   \\
2 & 151.2218766 & $-$7.7327286   & $20.796$ &         \\
3 & 151.3448200 & $-$7.7344961   & $20.919$ & $600$   \\
4 & 151.3199700 & $-$7.7700997   & $21.156$ &         \\
5 & 151.2891841 & $-$7.7723922   & $20.850$ &         \\
6 & 151.2874967 & $-$7.7732089   & $20.941$ & $900$   \\
7 & 151.3741900 & $-$7.6955633   & $21.454$ & $1800$  \\
8 & 151.3182500 & $-$7.7340325   & $21.209$ &         \\
9 & 151.2904600 & $-$7.7393686   & $21.669$ & $900$   \\
10 & 151.3438800 & $-$7.7470892  & $22.013$ &        \\
11 & 151.2734585 & $-$7.7342803  & $22.383$ & $1500$ \\
12 & 151.2570407 & $-$7.6970918  & $22.335$ & $1800$ \\
13 & 151.3160800 & $-$7.6916319  & $22.699$ &        \\
14 & 151.3259000 & $-$7.7242011  & $22.529$ &        \\
15 & 151.2894000 & $-$7.6638911  & $22.258$ &        \\
16 & 151.3644600 & $-$7.6930417  & $22.224$ &        \\
17 & 151.3693100 & $-$7.6728572  & $22.265$ & $1800$ \\
17a & 151.3694104 & $-$7.6731225 & $22.065$ &        \\
18 & 151.2543398 & $-$7.7673075  & $22.689$ &        \\
19 & 151.3036600 & $-$7.6752959  & $22.440$ &        \\
20 & 151.3209000 & $-$7.7312989  & $22.657$ &        \\
21 & 151.2540907 & $-$7.7510601  & $22.842$ &        \\
22 & 151.2999590 & $-$7.7639179  & $22.645$ &        \\
23 & 151.3402900 & $-$7.6925642  & $22.443$ &        \\
24 & 151.2907999 & $-$7.8078349  & $22.736$ &        \\
25 & 151.2916910 & $-$7.7087320  & $22.803$ &        \\
26 & 151.3211688 & $-$7.7692710  & $22.557$ &        \\
27 & 151.3629300 & $-$7.6982394  & $22.874$ & $2400$ \\
28 & 151.2869653 & $-$7.7839174  & $23.409$ &        \\
29 & 151.2487414 & $-$7.7048524  & $22.841$ &        \\
30 & 151.2889769 & $-$7.7638849  & $23.038$ &        \\ 
31 & 151.2377746 & $-$7.7306950  & $23.422$ &        \\
\hline \hline
\end{tabular}
\label{tab:UCD_Candidates_Table}
\end{table}

% ------------------ Observations Section -------------------- %
\subsection{Observations}
\label{sec:Observations}
We use the Keck Cosmic Web Imager (KCWI) \citep{Morrissey2018}, which is an integral field spectrograph on the $10\, \mathrm{m}$ Keck II telescope on Mauna Kea, to observe the UCD candidates described in Sec. \ref{sec:sample_selection} and shown in Fig. \ref{fig:GoodUCDs}.

The observations were conducted on the night of 2019 January 16 (Program ID: W142).
We used the Medium slicer with a Field of View (FoV) of $16.5\arcsec \times 20.4\arcsec$, $2\times2$ binning and low-resolution (BL) grating. We set the central wavelength to $4550\angstrom$. Overall, this set-up provides a nominal resolving power $R=1800$ and a wavelength coverage of $3500 < \lambda\, (\angstrom) < 5600$. The FoV of the Medium slicer also ensures that we have enough area around each UCD candidate to perform background subtraction.

At the end of the observing night, we took exposures of the standard star, Feige34, for the flux calibration.
The seeing conditions varied between $1\arcsec$ and $2\arcsec$ during the observations.
Table \ref{tab:UCD_Candidates_Table} also lists the UCDs observed and their total exposure times.

% ------------------ Data Reduction Section -------------------- %
\subsection{Data Reduction}
\label{sec:data_reduction}
We use the KCWI Data Extraction and Reduction Pipeline (KDERP, \url{https://github.com/Keck-DataReductionPipelines/KcwiDRP}) to reduce the raw data and obtain the final 3D datacubes.
The KDERP pipeline consists of eight data reduction stages, which are described in detail in the Github repository.

When reducing our raw data we skip the sky subtraction stage, as we prefer to perform a manual sky subtraction on the final reduced 3D datacubes due to the presence of the high background galaxy gradient in most of the datacubes.
We stack the reduced, non-sky subtracted datacube of each individual exposure for each UCD candidate, obtaining the final 3D cubes for each UCD.

% -------------------- UCD Radial Velocities and pPXF ---------------------- %
\section{Analysis}
\label{sec:data_analysis}
We visualize the final reduced UCD datacubes with the public available QFitsView\footnote{\url{http://www.mpe.mpg.de/~ott/QFitsView/}.} program and extract the spectrum of each object by taking a circular annulus around each UCD.
We choose an inner radius so as to maximize the signal coming from our object and an outer radius so as to best remove the sky$+$galaxy (i.e. NGC 3115) background contribution nearby the object.

The UCD results that we obtain in the following Sec. \ref{sec:kinematic_analysis}-\ref{sec:ucd_sizes} will be discussed in detail with their implications in Sec. \ref{sec:UCD_origin}.

\subsection{UCD radial velocities}
\label{sec:kinematic_analysis}
To measure the radial velocities of our observed UCDs, which are listed in Table \ref{tab:UCD_Candidates_Table}, we use the penalized pixel-fitting method (pPXF) initially developed by \citet{pPXF2004} and later implemented by \citet{pPXF2017}. 
We use the E-MILES \citep{Vazdekis2016} stellar library as template spectra at the constant spectral resolution of $\mathrm{FWHM} = 2.51\, \angstrom$. These stellar population synthesis models have ages up to $14\, \mathrm{Gyr}$ and metallicities of $-2.27 < \mathrm{[\mathit{Z}/H]} < +0.40$.
Because the resolution of the stellar library is very similar to that of the KCWI instrument ($\mathrm{FWHM}\sim2.53\, \angstrom$ at $\lambda = 4550\, \angstrom$), it is not necessary to convolve the spectral templates resolution to that of our observed spectra in pPXF.
This corresponds to $\sigma_\mathrm{inst}\sim71\, \mathrm{km\,s^{-1}}$ at the central wavelength.

We run pPXF with the default values of the multiplicative Legendre polynomials and bias parameters, while we use additive Legendre polynomials of the $10th$ order.

We perform the pPXF fit on the full wavelength range of $3600 < \lambda (\mathrm{\angstrom}) < 5550$, restricting it only in those cases where it would provide an improvement in the fit residuals. 
In fact, we find that excluding the bluer wavelengths, i.e. between $3600-3800\, \mathrm{\angstrom}$, provides better fit residuals for some UCDs. We adopt this narrower wavelength range in the pPXF fit, even though it does not significantly change the recovered radial velocity, which is still consistent within the uncertainties.

Table \ref{tab:Kinematics_UCDs} summarizes the radial velocities and 1-$\sigma$ uncertainties for each UCD candidate as recovered by pPXF. The radial velocities are converted to the heliocentric system, where the heliocentric velocity correction is calculated using the \textit{baryvel} function within the \texttt{PyAstronomy}\footnote{\url{https://github.com/sczesla/PyAstronomy}.} \citep{PyAstronomy} program, considering the location of the Keck telescope and the date of the observations. We calculate a heliocentric velocity correction $\Delta V_\mathrm{heliocorr} \sim 20\, \mathrm{km\,s^{-1}}$, which we add to the measured radial velocity of each UCD. 

\begin{table}
\caption{Summary table of the UCD radial velocities if measured, the fitting wavelength range and the S/N of the obtained spectra. Two of the objects (UCD 11 and 12) turned out to be background objects (see Appendix \ref{sec:background_objects}).}
\centering
\begin{tabular}{|c|c|c|c|c|c|c|}\hline\hline
UCD & $V_\mathrm{helio}$ ($\mathrm{km\,s^{-1}}$)   & $\lambda$ ($\mathrm{\angstrom}$) & S/N \\ \hline
1   & 450.0 $\pm$ 6.4     & $3700-5550$    & 26  \\
3   &  -                  & -              & 3   \\
6   & 538.0 $\pm$ 18.0    & $3800-5500$    & 13  \\
7   & 806.0 $\pm$ 24.0    & $3900-5500$    & 7   \\
9   & 599.0 $\pm$ 17.0    & $3900-5500$    & 6   \\ 
11  & 4856.0 $\pm$ 36.0   & $3600-5550$    & 3   \\
12  & 17550.0 $\pm$ 15.0  & $3600-5550$    & 3   \\
17  & 681.0 $\pm$ 20.0    & $3900-5550$    & 10  \\
17a & 1038.0 $\pm$ 14.0   & $3900-5550$    & 10  \\
27  & 772.0 $\pm$ 20.0    & $3900-5550$    & 7   \\
\hline \hline
\end{tabular}
\label{tab:Kinematics_UCDs}
\end{table}

Table \ref{tab:Kinematics_UCDs} also lists the UCD IDs with the wavelength range over which we perform the fit and the S/N of the observed object spectrum. For one object (UCD 3) we do not measure a radial velocity due to the low S/N.

We also measure the galaxy (i.e. NGC 3115) radial velocities where the background galaxy gradient is stronger. We recover a heliocentric radial velocity of $V_\mathrm{helio, 1} = 520.0(\pm 9.0)\, \mathrm{km\,s^{-1}}$ and $V_\mathrm{helio, 2} = 458.0(\pm 4.5)\, \mathrm{km\,s^{-1}}$ with S/N $=18$ at two locations in the FoV of the UCD 1 and UCD 9 datacubes, respectively. 
We confirm that these radial velocities are consistent with those of NGC 3115 at nearby positions by comparing them with the SLUGGS and MUSE \citep{Guerou2016} stellar kinematic measurements (see Fig. \ref{fig:2D_plot_stars}). 

\begin{figure*}
\centering
\includegraphics[width=1.\textwidth]{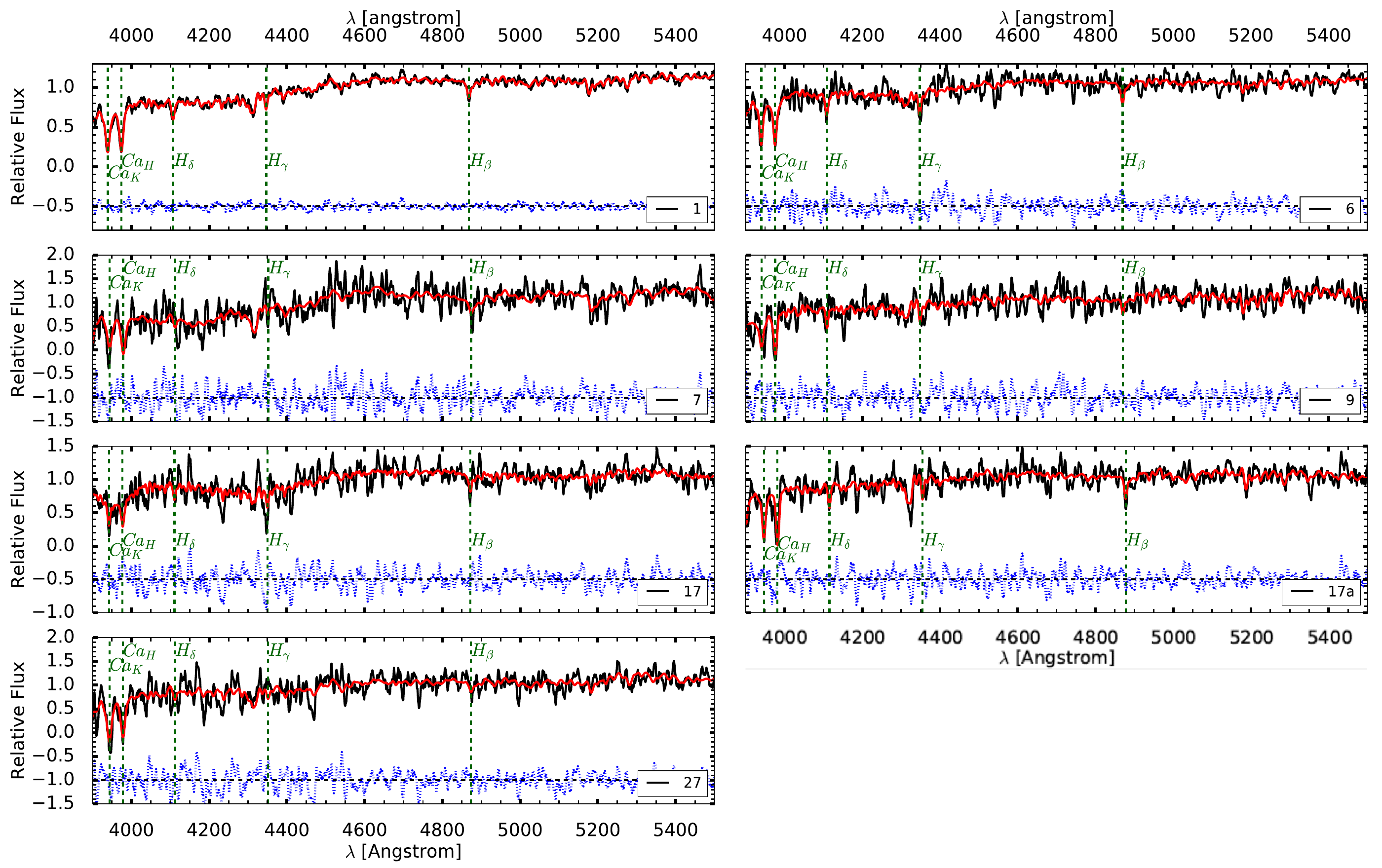}
\caption{pPXF best-fits (red line) to the observed UCD spectrum (black line). In blue are the fit residuals. The vertical dashed green lines label the five main absorption lines, which are fitted by pPXF and used to measure the radial velocity. The number on the bottom right of each plot indicates the UCD ID.}
\label{fig:pPXF_best-fits}
\end{figure*}

Fig. \ref{fig:pPXF_best-fits} shows the pPXF best fits to each UCD spectrum. The two candidates (UCD 11 and 12) that we identify as background objects from their radial velocity measurements are described in Appendix \ref{sec:background_objects}. In each plot of Fig. \ref{fig:pPXF_best-fits}, we label the main absorption lines. 

In summary, we obtain spectra for $9$ objects, out of the initial $14$ UCD candidates (see Fig. \ref{fig:GoodUCDs}), plus $1$ ``bonus" object using the KCWI instrument. From the spectral analysis, we conclude that $7$ are NGC 3115 members, having a radial velocity consistent with that of the galaxy, $2$ are background objects at higher redshifts and $1$ cannot be confirmed spectroscopically.
We therefore have radial velocities for a total of $12$ UCDs that are consistent with being members of NGC 3115, where $7$ come from our KCWI spectral analysis and $5$ come from previous work \citep{Pota2013}. In Fig. \ref{fig:AllGCsplusUCDs_1D}, we show the 2D spatial distribution of the $12$ UCDs on the plane of the sky around NGC 3115. 

% ------------------------- Stellar Population Section ----------------------- %
\subsection{Stellar Populations of UCD 1}
\label{sec:stellar_population}
Due to the low S/N of the observed UCDs, we perform a stellar population analysis only for UCD 1, which is the brightest of the confirmed UCDs with S/N $=26$.

We use a non-regularized parametrization, which provides a result that is more similar to the outcome of using the line-strength method, procuring single stellar population like star formation histories. We obtain mean mass-weighted estimates of $\mathrm{age}=(12\pm2)\, \mathrm{Gyr}$ and metallicity $\mathrm{[Z/H]}=-1.0\pm0.3$, indicating that UCD 1 is old and metal-poor. The errors are computed to account for the different results of using other methods such as the absorption line-strength analysis using the most commonly used pairs of line-indices, i.e. H$_{\beta}$ versus [MgFe] (e.g. \citealt{Thomas2003}).

% ------------------------- UCDs Size Section ----------------------- %
\subsection{UCD Sizes}
\label{sec:ucd_sizes}
We remeasure the sizes of all the $31$ UCD candidates identified in J14 by fitting a S\'ersic function \citep{Sersic1968} 
to each object within the QFitsView program. The final size of each object is given by calculating the average of the sizes measured separately in the $g$ and $z$ filters on the {\it HST}/ACS images. This method of measuring the sizes does not take into account the instrumental point-spread function (PSF), meaning that the sizes that we recover will be an overestimate of the true intrinsic size.
For the physical sizes of the objects we have assumed that they are located at the NGC 3115 distance (see Table \ref{tab:NGC3115_parameters}).

We generally measure similar sizes to J14, who measured the half-light radius of all their $31$ UCD candidates from the FWHM of a King \citep{King1962} profile fitted using the \textit{ishape} tool \citep{Larsen1999}. 
However for three objects (namely UCDs 3, 6 and 7), which are all part of our selected sample of UCD candidates shown in Fig. \ref{fig:GoodUCDs} and in two cases are spectroscopically confirmed members of NGC 3115 in Table \ref{tab:Kinematics_UCDs}, we measure much smaller sizes than J14.
We measure $\sim5\, \mathrm{pc}$ for UCDs 3 and 6, while J14 found them both to be $70.6\, \mathrm{pc}$. For UCD 7 we measure $\sim6\, \mathrm{pc}$, as opposed to the $73.6\, \mathrm{pc}$ of J14 (see also Appendix \ref{sec:issues} for a summary of the issues we encountered with the UCD candidates described in J14).

\begin{figure}
\centering
\includegraphics[width=0.5\textwidth]{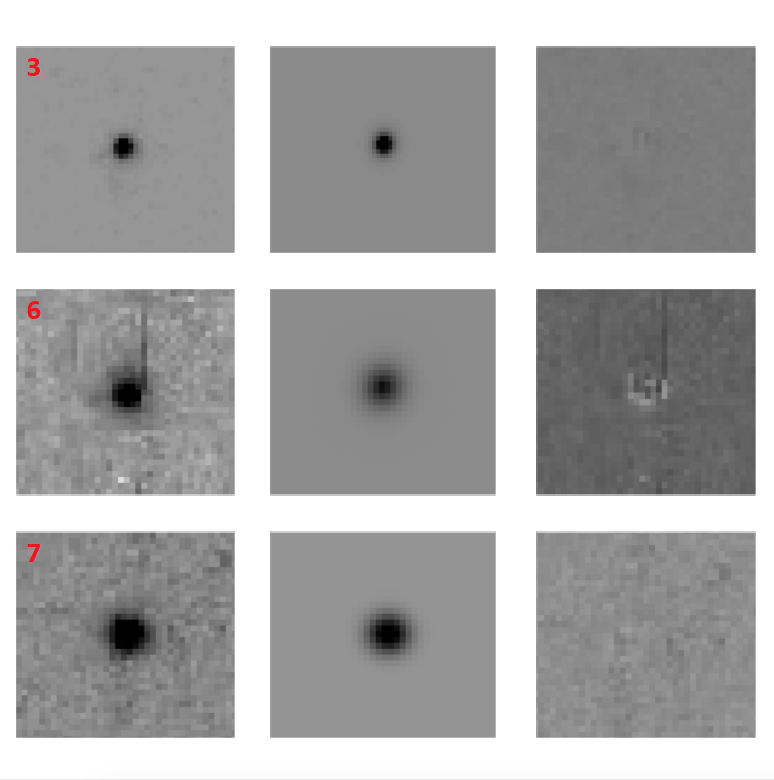}
\caption{{\it HST}/ACS images in the $g$ filter. From left to right we show the 2D image, model and residuals (i.e. $\mathrm{image}-\mathrm{model}$) from fitting a S\'ersic function to the three UCDs 3, 6, 7 for which we find the largest difference with respect to J14 size estimates. The residuals in the $z$ filter are very similar and give similar sizes. The number on the top left of the image indicates the UCD ID. The measured size of each UCD is summarised in Table \ref{tab:UCDs_Size_Table}.}
\label{fig:UCD367_QFitsViewFit}
\end{figure}

In Fig. \ref{fig:UCD367_QFitsViewFit}, we show as examples the model and residuals of the S\'ersic fits to the selected UCDs of Fig. \ref{fig:GoodUCDs}, i.e. UCD 3 (top), 6 (middle) and 7 (bottom), for which we find the largest discrepancy with the J14 size estimates.
From the 2D residuals, we notice that two UCDs are well represented by the fitted function, while the residuals are slightly worse in the case of UCD 6.

% -------------------- SKiMS, GCs, PNe Catalogs ---------------------- %
\section{NGC 3115 kinematic tracers}
\label{sec:TheGalaxy}

% ------------------- Description of the catalogs here --------------- %
\subsection{Stellar kinematics}
\label{sec:stellar_kinematic}

\begin{table}
\caption{Summary of NGC 3115 parameters. The values are taken from \citet{Brodie2014}, except for the stellar mass (from \citealt{Forbes2017a}).}
\centering
\begin{tabular}{|l|r|}\hline\hline
Distance ($d$)                                                    & $9.4\, \mathrm{Mpc}$      \\
Effective radius ($R_{e}$)                                        & $35\arcsec$               \\ 
Photometric axial-ratio ($q_\mathrm{phot}$)                       & $0.34$                    \\
Photometric position angle ($\mathrm{PA}_\mathrm{phot}$)          & $40\degr$                 \\
Systemic velocity ($V_\mathrm{sys}$)                              & $663\, \mathrm{km\,s^{-1}}$ \\
Stellar mass ($\log{(\mathrm{M_{*}}/\mathrm{M_{\odot}})}$)        & $10.93$ \\
\hline \hline
\end{tabular}
\label{tab:NGC3115_parameters}
\end{table}

The stellar kinematics of NGC 3115 used in this work come from \citet{Arnold2011,Arnold2014} and were obtained using the Stellar Kinematics from Multiple Slits (SKiMS) technique \citep{Proctor2009,Foster2009,Foster2011}. 
These data, obtained with the Keck/DEIMOS instrument \citep{DEIMOS2003}, contain
$166$ measurements of not only the galaxy kinematics with their locations (RA and DEC), but also a quality flag. 
We use, therefore, only the stellar kinematic data deemed as good, i.e. flag=A.
This way, we have a final catalogue of $126$ stellar kinematic measurements with radial velocities over the range $400 < V (\, \mathrm{km\,s^{-1}}) < 1000$.
We complement the SLUGGS stellar kinematics data with the MUSE data from \citet{Guerou2016}, publicly available in the form of 2D spatially resolved maps. 

\begin{figure}
    \centering
    \includegraphics[width=0.5\textwidth]{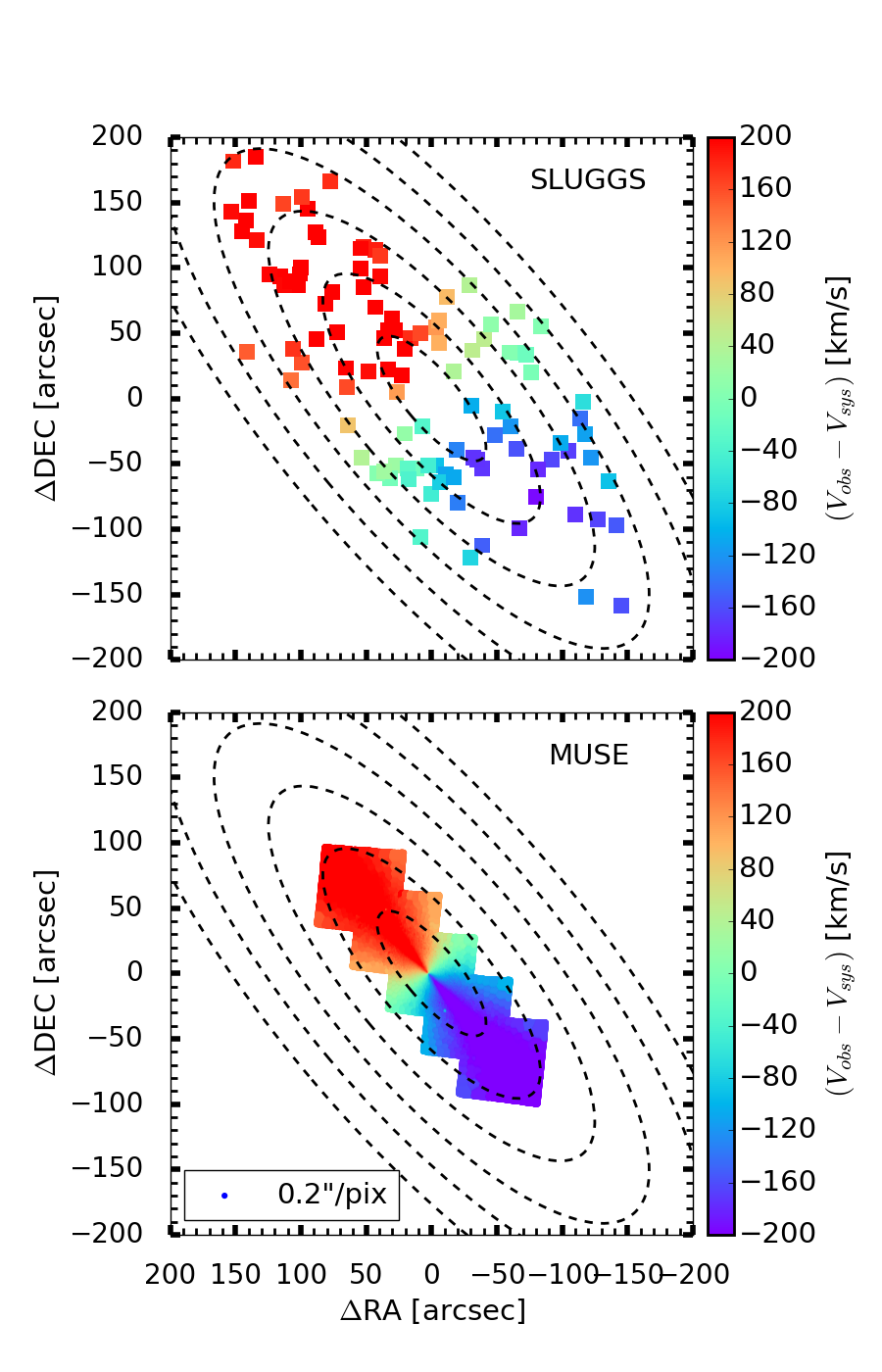}
    \caption{2D sky-projected maps of the SLUGGS (top) and MUSE (bottom; \citealt{Guerou2016}) stellar kinematics data. The data-points on the maps are colour-coded according to the velocity scale shown by the colour-bar on the right-hand side. The black dashed lines represent the $1-6\, R_\mathrm{e}$ photometric ellipses, inclined by the $\mathrm{PA}_\mathrm{phot}\sim40\degr$ (Table \ref{tab:NGC3115_parameters}) of the galaxy.}
    \label{fig:2D_plot_stars}
\end{figure}

Fig. \ref{fig:2D_plot_stars} shows the SLUGGS and MUSE \citep{Guerou2016} stellar kinematics data of NGC 3115 as 2D sky-projected maps, colour-coded using the same velocity scale.
We represent the $1-6\, R_\mathrm{e}$ photometric ellipses, rotated by the $\mathrm{PA}_\mathrm{phot}$ and flattened by the photometric axial-ratio $q_\mathrm{phot}$ of the galaxy, as black dashed lines (see Table \ref{tab:NGC3115_parameters} for the values of the NGC 3115 photometric parameters). 
We emphasize the larger extent of the SLUGGS stellar kinematics data (which extend out to $\sim4\, R_\mathrm{e}$), as compared with the \citep{Guerou2016} data. The latter, in fact, extend out to only $\sim2\, R_\mathrm{e}$ but do probe the central regions of NGC 3115 with a pixel scale of $0.2\arcsec/\mathrm{pixel}$, unlike SLUGGS that probes only beyond $\sim1\, R_\mathrm{e}$.
Additionally, we see that the stellar system displays good symmetry with respect to the galaxy major- and minor-axes and that it is strongly rotating around the minor-axis, i.e. $\mathrm{PA}_\mathrm{kin}\simeq\mathrm{PA}_\mathrm{phot}$. 

\subsection{The Globular Cluster System}
\label{sec:globular_clusters}
NGC 3115 has a large number of GCs, $150$ of which have been confirmed with spectroscopic analyses \citep{Arnold2011,Pota2013,Forbes2017}. 
This GC catalogue, observed as part of the SLUGGS survey, contains the radial velocities with associated uncertainties and locations for each object, and it has been cleaned to exclude foreground star contaminants \citep{Arnold2011}. 
Photometry is available for all the GCs with $g$, $r$ and $i$ magnitudes obtained from Subaru/Suprime-Cam images \citep{Arnold2011,Pota2013}. 
Additionally, as part of SLUGGS, J14 measured the half-light radii, $g$ and $z$ magnitudes for $360$ GC candidates to which they added $421$ additional GC candidates from Suprime-Cam \citep{Arnold2011}.
This led to a final catalogue of $781$ GC candidates, $176$ of which were spectroscopically confirmed \citep{Arnold2011,Pota2013}.

In our analysis, we use a catalogue of $191$ unique GCs around NGC 3115 by combining and matching the individual spectroscopic catalogs from J14 and \citet{Forbes2017}. 
This final catalogue has a radial velocity measurement for each GC in the range $350 < V\, \mathrm{(km\,s^{-1})} < 1200$, where the lower limit excludes the foreground stars. 
$145$ of these velocity measurements have associated uncertainties of the order of $5-10\, \mathrm{km\,s^{-1}}$. The remaining $46$ GCs do not have quoted uncertainties. For these objects, we assume an average uncertainty of $20\, \mathrm{km\,s^{-1}}$.

\begin{figure}
    \includegraphics[width=0.5\textwidth]{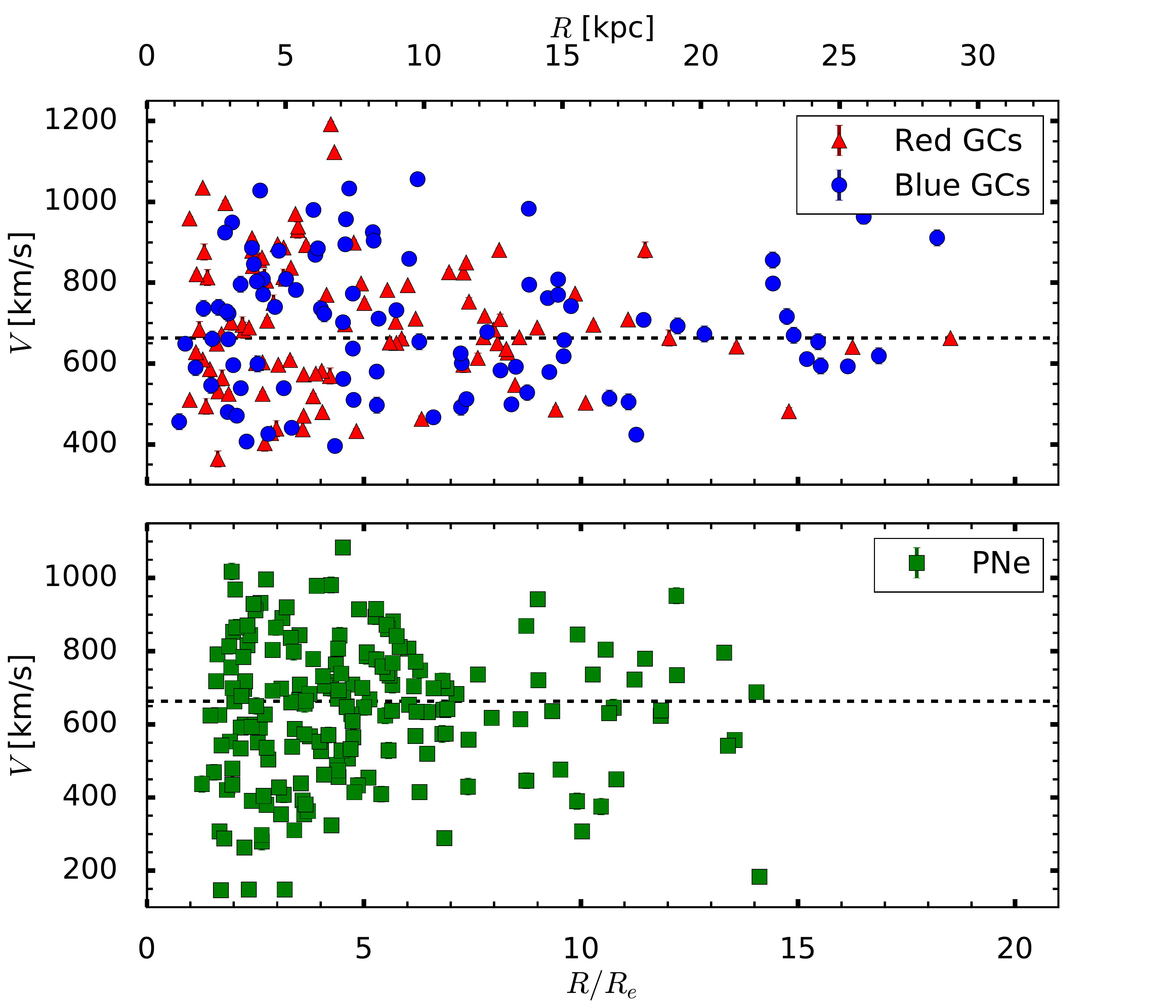}
    \caption{Top panel: the blue and red GC sub-populations of NGC 3115, separated by adopting the colour-split $(g-i)=0.93$ \citep{Brodie2012}.
    The radial velocity is plotted as a function of the galactocentric radius, $R/R_{e}$, and the horizontal black dashed line indicates the systemic velocity, $V_\mathrm{sys}\sim663\, \mathrm{km\,s^{-1}}$ (Table \ref{tab:NGC3115_parameters}), of the galaxy. 
    Bottom panel: the PNe system of NGC 3115.}
    \label{fig:GCs_PNe_Vobs_vs_Radius}
\end{figure}

Fig. \ref{fig:GCs_PNe_Vobs_vs_Radius} shows the observed velocities of the final GC catalogue plotted as a function of the galactocentric radius, $R/R_{e}$, where for the galactocentric radius, $R$, of each object we have adopted the equivalent circular radius defined as 

\begin{equation}
    R_\mathrm{circ} = \sqrt{ q_\mathrm{phot}(\Delta X)^{2} + (\Delta Y)^{2}/q_\mathrm{phot} }.
    \label{eq:circ_radius}
\end{equation}

In the above Eq. \ref{eq:circ_radius}, $\Delta X$ and $\Delta Y$ are the Cartesian coordinates of each object measured from the galaxy centre along its photometric major- and minor-axes, respectively. The $q_\mathrm{phot}$ parameter is defined in Table \ref{tab:NGC3115_parameters}.

We divide the GC sample into two sub-populations by implementing a colour-split $(g - i) = 0.93$ \citep{Brodie2012} to distinguish between the red, metal-rich ($100$ objects, red triangles in Fig. \ref{fig:GCs_PNe_Vobs_vs_Radius}) and the blue, metal-poor ($91$ objects, blue dots) GC sub-populations. 

\subsection{The Planetary Nebulae System}
\label{sec:planetary_nebulae}
The PNe catalogue of NGC 3115 used in this work comes from the ePN.S survey of ETGs \citep{Pulsoni2018Survey}. This is currently one of the largest extra-galactic PNe surveys, obtaining kinematic measurements of large samples of PNe in $33$ early-type galaxies. 
Because PNe are strong [O~{\sc III}] $5007\, \mathrm{\angstrom}$ emitters, they can be observed far out from the galaxy center in regions of low surface brightness. Additionally, they have been shown to be good kinematic tracers of the stellar component in the host galaxy, and therefore can be used as probes of the properties of the underlying stellar population in the halos of ETGs (e.g. \citealt{Romanowsky2006,Hartke2017}).
The PNe catalogue of NGC 3115 contains $189$ objects with radial velocities over the range $140 < V\, \mathrm{(km\,s^{-1})} < 1100$ and mean uncertainty of $\delta V = 20\, \mathrm{km\,s^{-1}}$ \citep{Pulsoni2018Survey}. We do not apply any velocity cuts to the catalogue. 
The surface density profile of the PNe of NGC 3115 has been shown to agree well with the stellar surface brightness profile \citep{Cortesi2013a}, which means that we can effectively use PNe as proxies of the stars at large radii in NGC 3115.  
Fig. \ref{fig:GCs_PNe_Vobs_vs_Radius} shows the observed velocities of the PNe system plotted as a function of the galactocentric radius, $R/R_{e}$, with $R$ defined as in Eq. \ref{eq:circ_radius}.

% ------------------- Point-Symmetry Tests ----------------- %
\section{Symmetry Tests}
\label{sec:PointSymmetry}
We can produce a denser distribution of tracers by adopting the \textit{folding} method, similarly to \citet{Peng2004,Cappellari2015,Pulsoni2018Survey}. 
By doing this, we can improve the spatial coverage of the kinematic maps that we generate from an underlying tracer distribution.
However, before \textit{folding} each catalogue we need to test whether our kinematic data are actually symmetric.

In this section we analyse the symmetry of the GC and PNe systems of NGC 3115.
In order to test the symmetry, we adopt a similar procedure to that used in \citet{Coccato2013} and \citet{Pulsoni2018Survey}. 
We compare the velocities of the tracers on one side of the galaxy with those on the other side of the galaxy with respect to the photometric major-axis.
By definition, an axi-symmetric system would have velocities, $V(R, \Phi)$, at galactocentric radius, $R$, and angle, $\Phi$, for $0\degr < \Phi < 180\degr$, equal to the velocities, $V(R, -\Phi)$, at galactocentric radius $R$ and angle $(-\Phi)$ with respect to the photometric major-axis of the galaxy.

In our analysis, we have first rotated our galaxy coordinates such that the photometric major-axis lies on the $x$-axis of an orthogonal reference system centered on NGC 3115 and oriented $y$ toward North and positive $x$ toward East. 
The new rotated coordinates, $\Delta X$ and $\Delta Y$, of each point are given by

\begin{equation}
\begin{split}
    \Delta X &= (\mathrm{RA} - \mathrm{RA}_{0})\cos{\theta} - (\mathrm{DEC} - \mathrm{DEC}_{0})\sin{\theta} \\
    \Delta Y &= (\mathrm{RA} - \mathrm{RA}_{0})\sin{\theta} + (\mathrm{DEC} - \mathrm{DEC}_{0})\cos{\theta},
    \end{split}
\label{eq:CoordinateRotation}
\end{equation}

where RA, DEC are the original coordinates of each point in our catalogs and $\mathrm{RA}_{0}$, $\mathrm{DEC}_{0}$ are those of the centre of NGC 3115. 
$\theta$ is the rotation angle defined as $\theta = (\mathrm{PA}_\mathrm{phot}-90\degr)$.

In this new system, we then compare the velocities of our tracers with $0\degr < \Phi < 180\degr$, where $\Phi$ is the angular separation of the object from the major-axis, i.e. $x$-axis of the reference system, and it is measured anti-clockwise, with that of the tracers with $180\degr < \Phi < 360\degr$. The more these two velocity fields overlap, the more the system is consistent with being axi-symmetric. Any significant difference between these two velocity fields would suggest deviations from axi-symmetry that could be due, for instance, to recent galaxy interactions that perturbed the system from its dynamical equilibrium state.

\begin{figure*}\centering
    \includegraphics[width=0.72\textwidth]{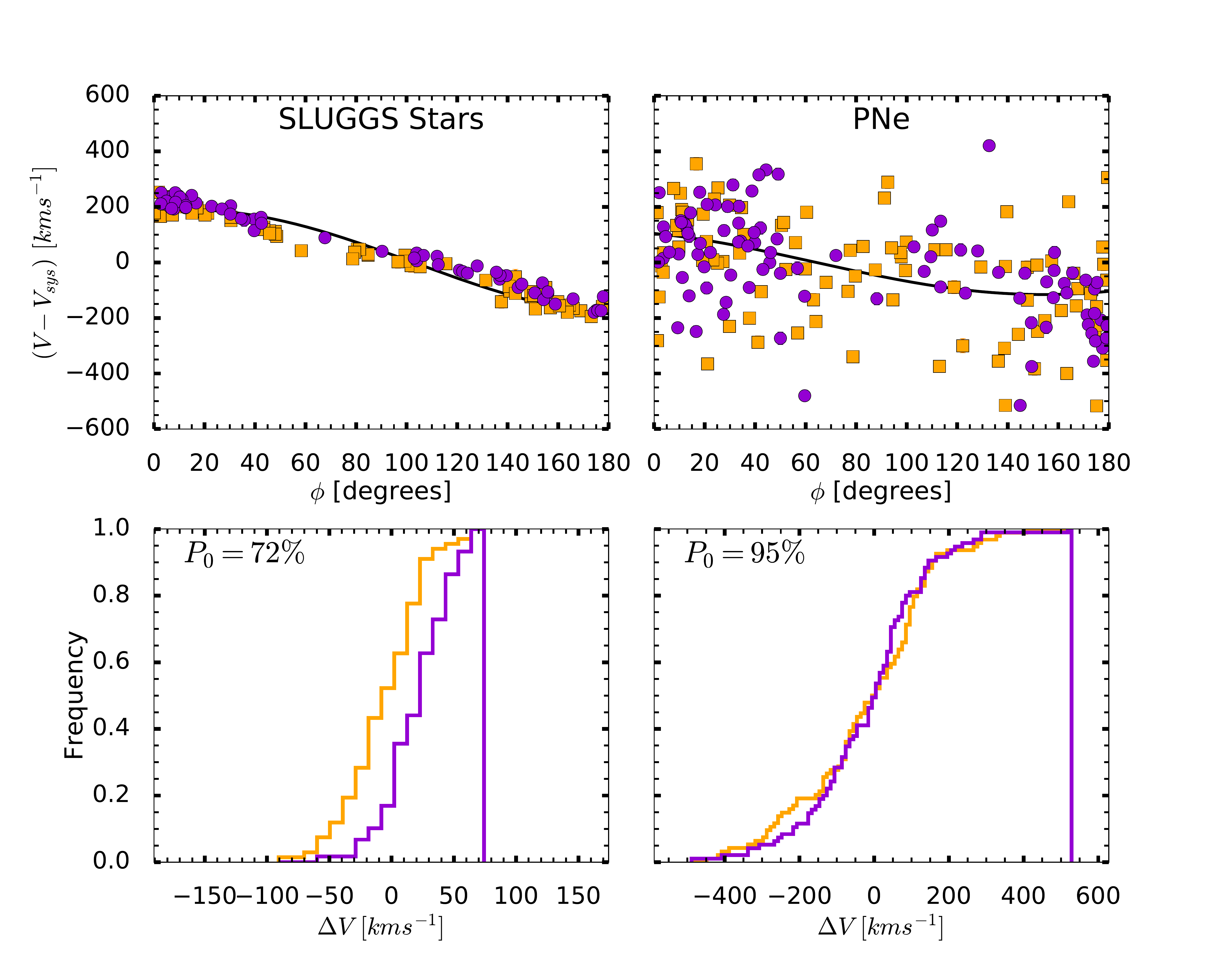}
    \includegraphics[width=0.72\textwidth]{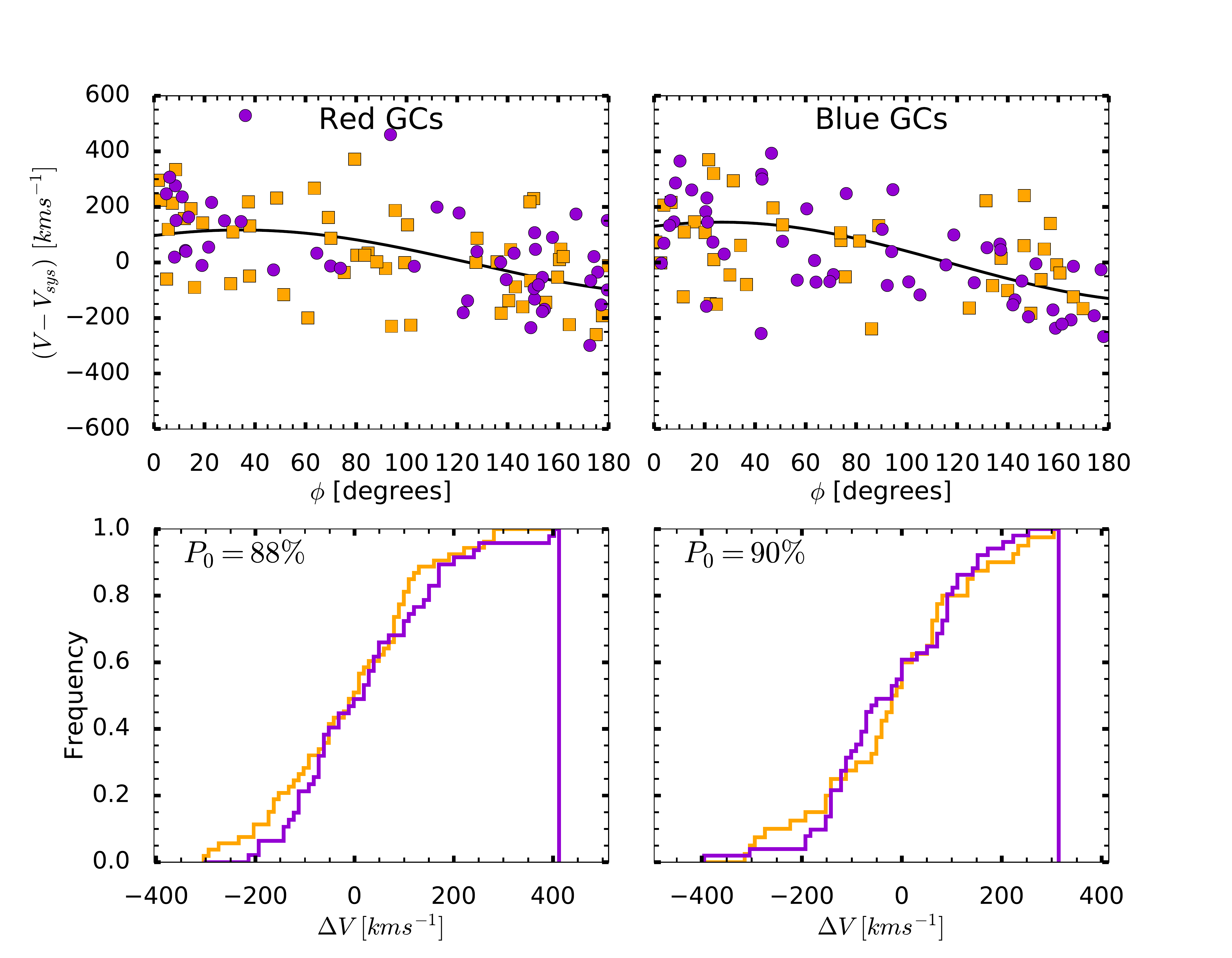}
    \caption{Symmetry tests. \textit{Top panels:} velocities of our kinematic catalogs plotted as a function of the position angle, $\Phi$, of each tracer for all radii. We compare the velocities of the tracers on one side of the galaxy, $V(0\degr < \Phi < 180\degr)$ (yellow squares), with those on the other side of the galaxy, $V(180\degr < \Phi < 360\degr)$ (magenta circles), folded with respect to the photometric major-axis of NGC 3115. The black curve is the best-fit rotation model curve to all our kinematic data defined in Eq. \ref{eq:cosine_law}. \textit{Bottom panels:} measured value of the difference, $\Delta V\, \mathrm{(km\,s^{-1})}$, between the individual velocities of our tracers at their position angle, $\Phi$, from the fitted rotation model curve defined in Eq. \ref{eq:cosine_law}. The corresponding histograms are shown in the form of cumulative density functions, normalized to one, with a bin width of $10$. On the top left of each panel, we indicate the probability value, $P_{0}$, for the null hypothesis that the two functions are drawn from the same distribution is true.}
    \label{fig:point_symmetry_catalogs}
\end{figure*}

Fig. \ref{fig:point_symmetry_catalogs} shows the results of our axi-symmetry tests performed on all the catalogs described in the previous Sections \ref{sec:stellar_kinematic}--\ref{sec:planetary_nebulae}.
Here, we plot the velocity of the tracers, $V(0\degr < \Phi < 180\degr)$ (yellow squares) and that of the tracers $V(180\degr < \Phi < 360\degr)$ (magenta circles), folded with respect to the photometric major-axis of the galaxy, at all radii as a function of the angular separation, $\Phi$, expressed in degrees.
The black solid line is the best-fit rotation model curve that we obtain by fitting the cosine law function, given by Eq. \ref{eq:cosine_law} \citep{Proctor2009,Foster2016,Bellstedt2017}

\begin{equation}
V_\mathrm{obs} = V_\mathrm{sys} \pm V_\mathrm{rot}\cos{\Phi}, 
\label{eq:cosine_law}
\end{equation}

where the angular separation, $\Phi$, is defined by Eq. \ref{eq:cosine_law2}

\begin{equation}
	\tan{\Phi} = \frac{\tan{(\mathrm{PA} - \mathrm{PA}_\mathrm{kin})}}{q_{\mathrm{kin}}}
\label{eq:cosine_law2}
\end{equation}

In Eq. \ref{eq:cosine_law2}, $\mathrm{PA}$ is the position angle of the object, $\mathrm{PA}_{\mathrm{kin}}$ is the kinematic position angle of the galaxy and $\mathrm{q}_{\mathrm{kin}}$ is the kinematic axial-ratio of the galaxy.
For only these symmetry tests, following \citet{Coccato2013}, we do not consider the kinematic axial-ratio, which we simply set $\mathrm{q}_{\mathrm{kin}}=1$ in Eq. \ref{eq:cosine_law2}, when fitting for the cosine law function to our data. In this way, Eq. \ref{eq:cosine_law} reduces to 

\begin{equation}
	V_\mathrm{obs} = V_\mathrm{sys} \pm V_\mathrm{rot}\cos{(\phi - \mathrm{PA}_\mathrm{kin})},
\label{eq:cosine_law3}
\end{equation}

where $PA_\mathrm{kin}$ and $V_{\mathrm{rot}}$ are the free kinematic parameters in the fit and $\phi$ is the $x$ variable between $[0,180]\degr$.
In fitting Eq. \ref{eq:cosine_law3} to our data, we have already subtracted the systemic velocity, $V_\mathrm{sys}\sim663\, \mathrm{km\,s^{-1}}$ of NGC 3115 from the observed radial velocity of all our tracers. 
We highlight that in Sec. \ref{sec:kinemetry}, we do take the kinematic axial-ratio parameter, $\mathrm{q}_{\mathrm{kin}}$, into account in the full modelling process. 

For each one of our kinematic catalogs, we measure the difference between the velocity of each tracer at its position angle $\phi$ from the best-fit model curve of Eq. \ref{eq:cosine_law3}. 
In the form of histograms, we plot the measured velocity difference of the tracers with $0\degr < \phi < 180\degr$ (yellow line) and that of the tracers with $180\degr < \phi < 360\degr$ (magenta line). The histograms are given as cumulative density distributions, normalized to one, and such that each bin has a width of $\Delta V = 10\, \mathrm{km\,s^{-1}}$. 

When we compare the velocity scatters from the best-fit model rotation curve of our tracers, on the two different sides of the galaxy with respect to its photometric major-axis, we observe that the two distribution functions are similar for all four datasets. Additionally, from the velocity as a function of the position angle plot, we observe that the datasets on either side of the photometric major-axis, display a good overlap. 
This means that the two distribution functions are likely tracing a similar underlying spatial distribution on both sides of the galaxy.  
We point out that the observed small deviations of the SLUGGS stars from the best-fit model rotation curve are taken into account when including higher order moments in Eq. \ref{eq:cosine_law3}. 
In fact, if we include harmonics of the third order in Eq. \ref{eq:cosine_law3} for the fitting of our kinematic data (as in equation 10 of \citealt{Pulsoni2018Survey}), then the small deviations of the data from the best-fit model curve are corrected for, yielding an improvement of the fit also seen in the corresponding histogram of the velocity difference. However, the results are overall consistent in the case with and without higher order harmonics and suggest that the SLUGGS stars have symmetric properties, as shown also by the good overlap between the data on the two sides of the galaxy with respect to the photometric major-axis of NGC 3115 (see Table \ref{tab:NGC3115_parameters}).

We perform the non-parametric Kolmogorow--Smirnov (KS) test to quantify whether the two galaxy sides are indeed symmetric with respect to the photometric major-axis.
We run the two-sample KS test on the cumulative distribution functions of the velocity scatters from the best-fit model curve of the two different sides of the galaxy, obtaining that the two distributions are similar with probabilities of $P_{0}=72\%$, $95\%$, $88\%$, $90\%$ for the SLUGGS stars, PNe, red and blue GCs, respectively. In Fig. \ref{fig:point_symmetry_catalogs}, we indicate the recovered $P_{0}$ for all four datasets. 
The results from the KS statistic mean that we cannot reject the null hypothesis that the two functions are drawn from the same distribution at the chosen confidence interval, suggesting that the velocity distributions of our kinematic datasets with respect to the photometric major-axis of NGC 3115 are similar.

Our results for the PNe are also in agreement with the previous results from \citet{Pulsoni2018Survey}, who have found that the PNe system of NGC 3115 is point-symmetric. 

Because our tests indicate that the NGC 3115 kinematics is axi-symmetric with respect to its photometric major-axis, we can apply the \textit{folding} method to each of our catalogues.
We take the original tracer distribution, $(X, Y, V)$, of each catalogue and create its symmetric counterpart, $(X, -Y, V)$. 
The final sample then contains the original tracer distribution plus its symmetric one. We end up with a final catalogue twice as large as the original to be used in the following section.

% ------------------- Kriging Section ---------------------- %
\section{Results}
\label{sec:results}

\subsection{2D Kinematic Maps}
\label{sec:2D_kinematic_maps}
We use the kriging spatial interpolation technique \citep{Krige1951} to produce 2D continuous kinematic maps of our discrete tracers in order to visualize their 2D kinematics in a similar manner to that of the stars. Thus we adopt a different approach than \citet{Pota2013a}, who only produced 1D profiles of GC kinematics.
In astronomy kriging is a powerful tool because the type of interpolation is inverse-noise-weighted meaning that low S/N data will have a reduced impact on the kinematic map.
Here we use the kriging method as implemented by \citet{Pastorello2014}.
This method has been applied in previous works to produce 2D maps of the kinematic moments of galaxies (e.g. \citealt{Proctor2009,Foster2013,Foster2016,Bellstedt2017}).

We produce the kriging maps for all our kinematic datasets, described in the previous Secs. \ref{sec:stellar_kinematic}-\ref{sec:planetary_nebulae}, after \textit{folding} each dataset with respect to the photometric major-axis of the galaxy. By doing this, we aim to improve the spatial resolution of the maps. This \textit{folding} is  important for the discrete tracers, i.e. GCs and PNe, in order to obtain a better velocity dispersion measurement from them. This is less important for the stellar kinematics as their velocity dispersion is an integrated line-of-sight quantity, (unlike GCs and PNe which are discrete tracers).

For the GCs and PNe, we spatially bin our \textit{folded} catalogues such that each bin contains the $15$ nearest objects. From this we calculate the average velocity and the velocity dispersion (from the standard deviation of the tracer velocities from the mean) in each bin. 
These averaged velocity and velocity dispersion measurements, in each bin, are used to generate the 2D kinematic maps described below.

\subsubsection{SLUGGS integrated starlight}
The left-hand side of Fig. \ref{fig:2D_KrigingMaps} (first row) shows that the velocity of the stars of NGC 3115 is characterized by a strong rotational signature which is well aligned with the overall photometric major-axis of the galaxy out to $\sim4\, R_\mathrm{e}$. Previous works had already highlighted the rotationally supported feature of NGC 3115, characterized by a steep rise of the line-of-sight velocity to $V_{rot}\sim260\, \mathrm{kms^{-1}}$ and by a subsequent flattening within the inner $\sim2\, R_{\mathrm{e}}$ from the galaxy centre \citep{Rubin1980,Capaccioli1993,Kormendy1996}. 
In Fig. \ref{fig:2D_KrigingMaps}, the rotation velocity rises very steeply to values $\sim200\, \mathrm{km\,s^{-1}}$, at $\sim1\, R_\mathrm{e}$, and then it remains roughly constant with some weak evidence of decrease beyond $\sim2\, R_\mathrm{e}$. Along the minor-axis, the velocity map does not show any signs of significant rotation. 
The stellar velocity dispersion map on the right-hand side of Fig. \ref{fig:2D_KrigingMaps} shows a higher dispersion, $\sim150\, \mathrm{km\,s^{-1}}$, within the central regions of the galaxy that extends along the photometric minor-axis. Along the photometric major-axis, the velocity dispersion is lower, correspondingly to the regions of high rotation amplitude in the velocity map, with values of $\sim100\, \mathrm{km\,s^{-1}}$ beyond $\sim1\, R_\mathrm{e}$. The velocity dispersion remains then roughly flat along the entire major-axis.

\subsubsection{PNe}
The PNe tracers also show a rise in rotation velocity at $\sim2\, R_\mathrm{e}$ along the major-axis (second row of Fig. \ref{fig:2D_KrigingMaps}, left-hand side). Here the velocity reaches a maximum of $\sim180\, \mathrm{km\,s^{-1}}$ followed by a decrease beyond $2\, R_\mathrm{e}$ and out to $\sim5-6\, R_\mathrm{e}$ where it flattens to lower velocity values down to $\sim60\, \mathrm{km\,s^{-1}}$. 
The velocity dispersion map of the PNe (on the right-hand side of Fig. \ref{fig:2D_KrigingMaps}) is overall constant at $\sim\, \mathrm{150\, \mathrm{km\,s^{-1}}}$. There is a region of higher velocity dispersion that is offset from the major-axis between $\sim3-4\, R_\mathrm{e}$. However, this does not correspond with any peculiarity in the velocity map, hence it could simply be an artificial feature due to the folding technique and not represent any real feature. 

\subsubsection{GCs}
The third and fourth rows of Fig. \ref{fig:2D_KrigingMaps} show the 2D kinematic maps of the red and blue GCs, respectively.
From the 2D velocity maps, we notice that both GC sub-populations show a significant amount of rotation along the major-axis with a steep rise in velocity to $\sim180-200\, \mathrm{km\,s^{-1}}$, followed by a decrease to $\sim100\, \mathrm{km\,s^{-1}}$ at larger radii. However, the red and blue GC velocity maps differ from each other as the former peaks between $\sim1$ and $\sim2\, R_\mathrm{e}$, while the latter peaks at larger radii between $\sim2$ and $\sim3\, R_\mathrm{e}$.
The velocity dispersion maps show colder "spots" in the corresponding regions of high rotation velocity, but are overally constant in the rest of the maps.

\begin{figure*}
    \includegraphics[width=0.48\textwidth]{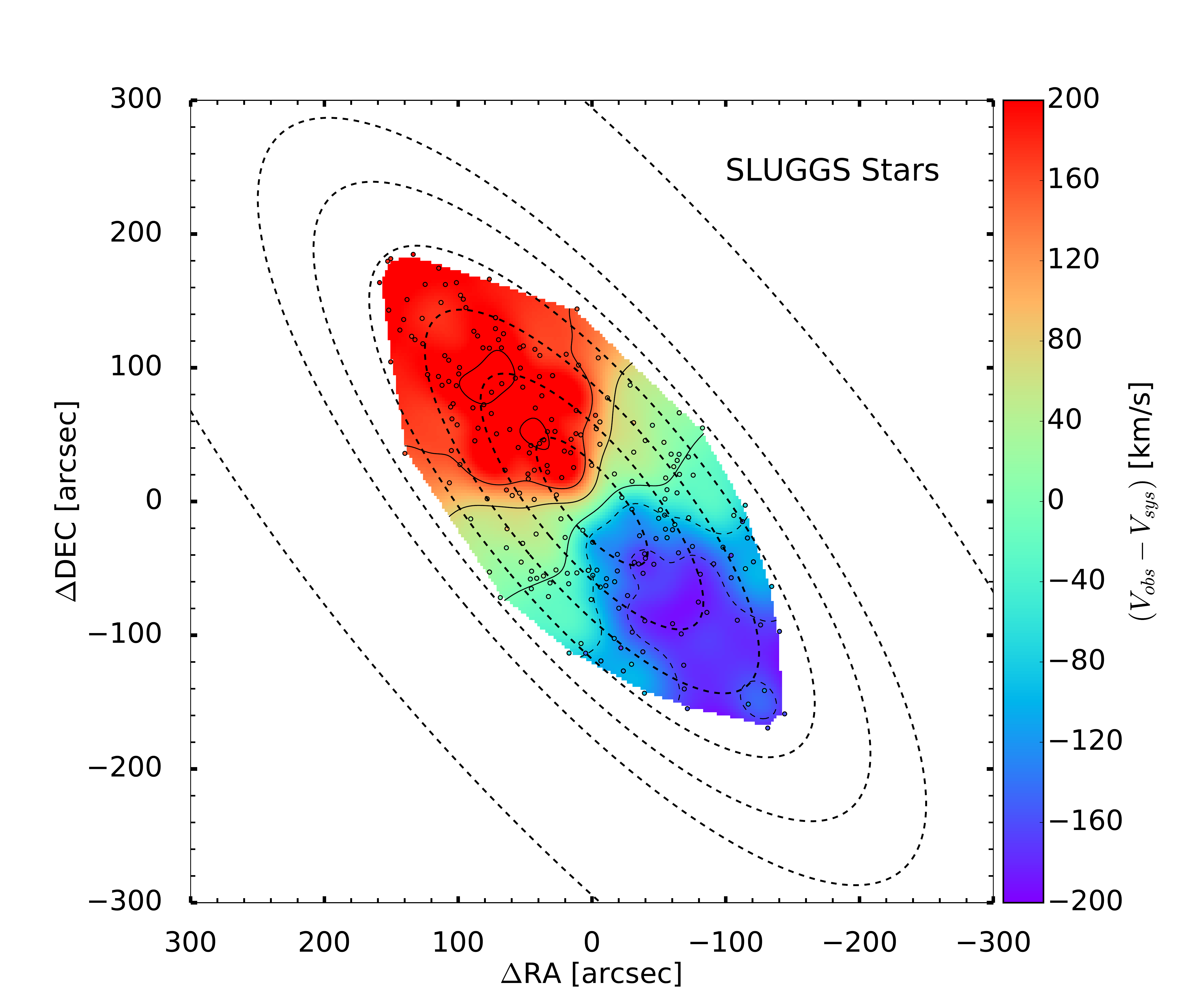}
    \includegraphics[width=0.48\textwidth]{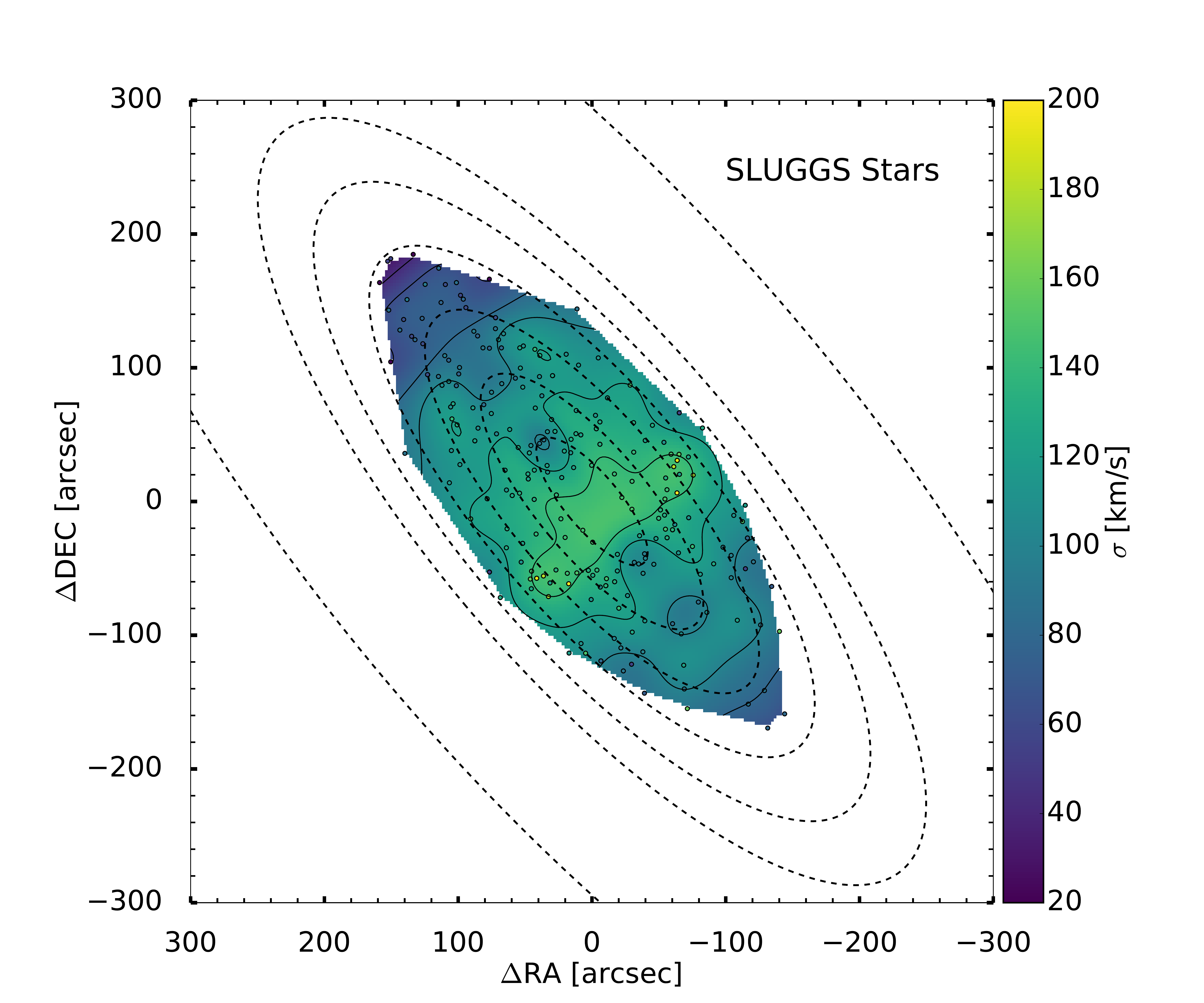}
    \includegraphics[width=0.48\textwidth]{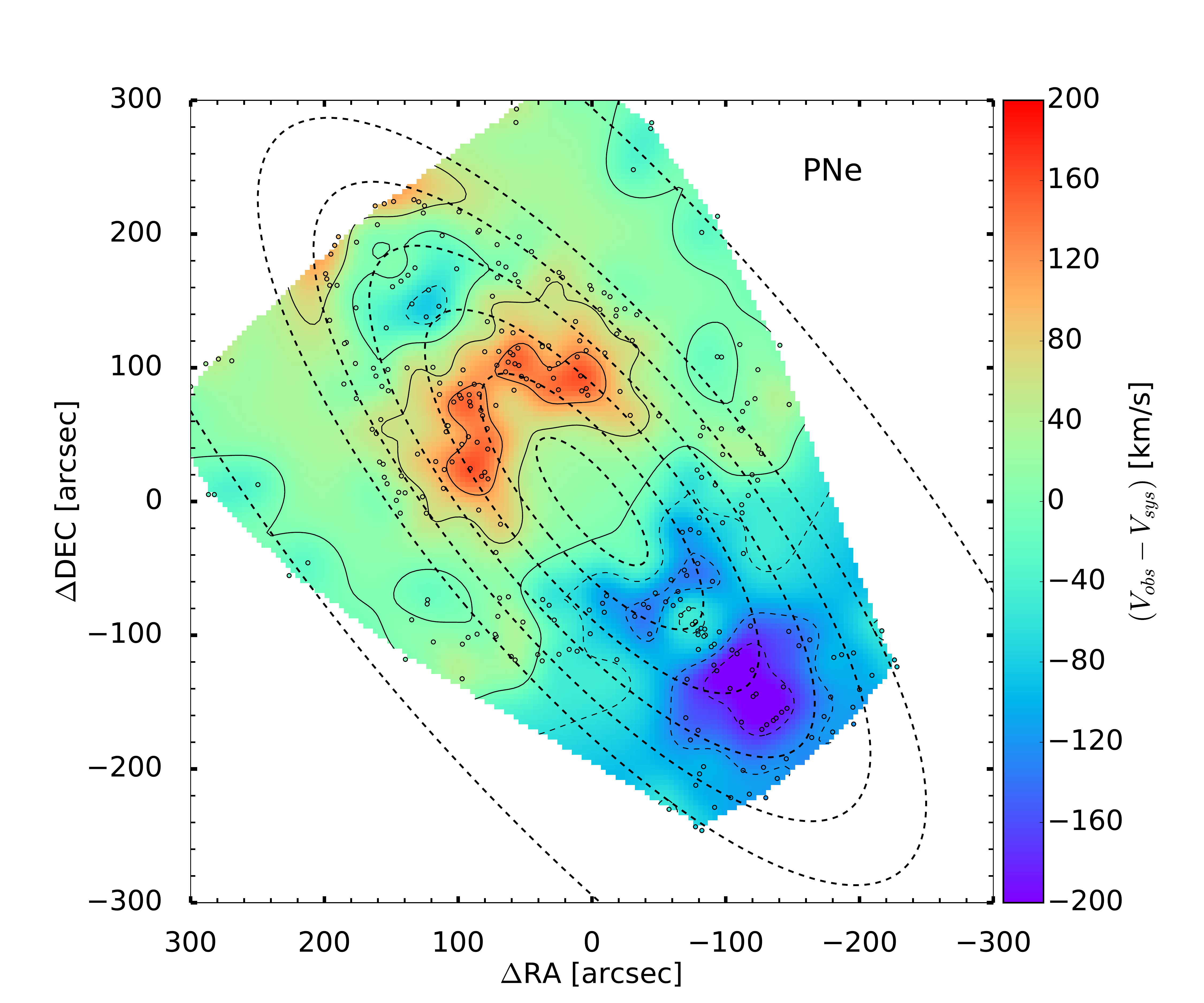}
    \includegraphics[width=0.48\textwidth]{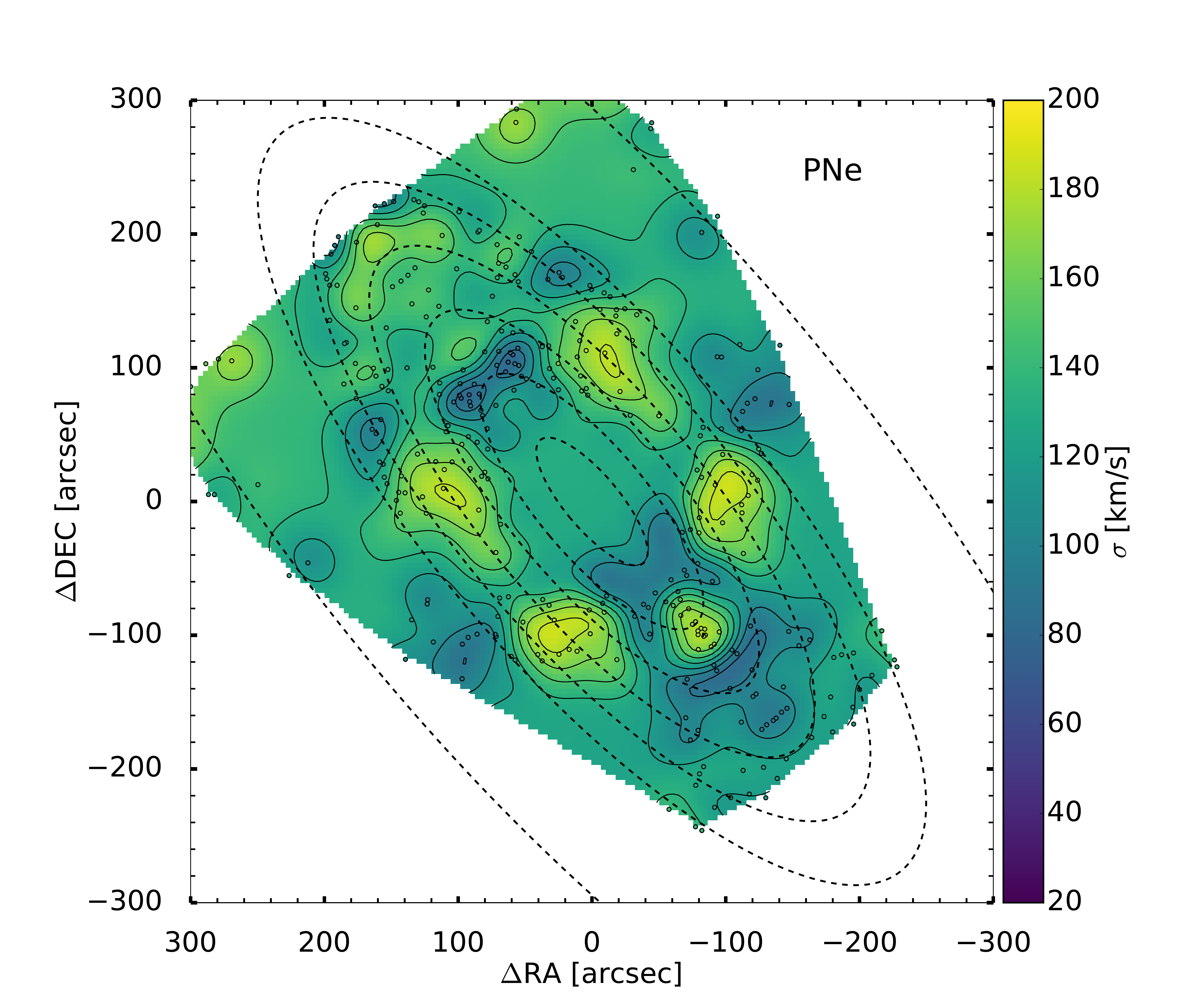}
    \includegraphics[width=0.48\textwidth]{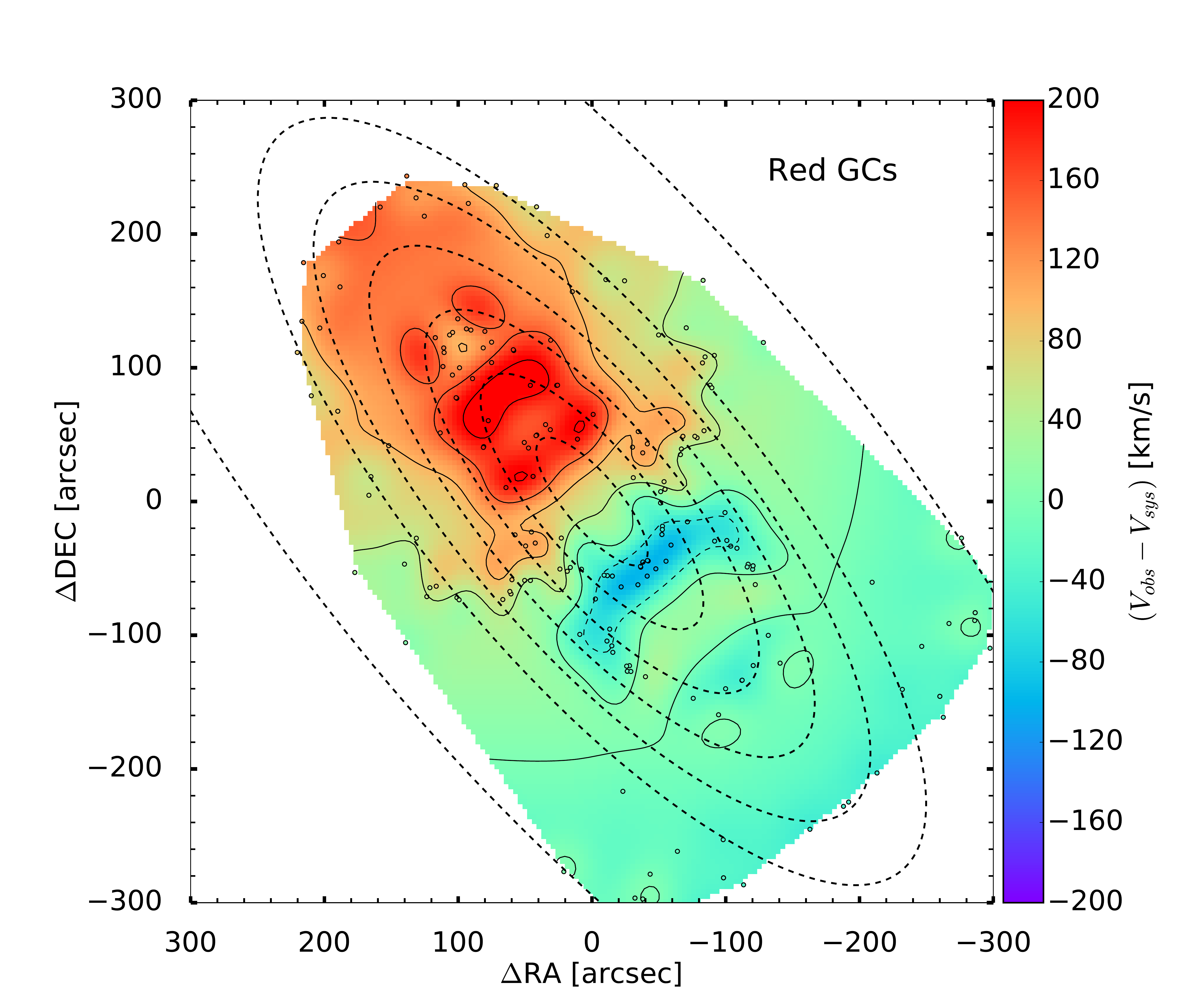}
    \includegraphics[width=0.48\textwidth]{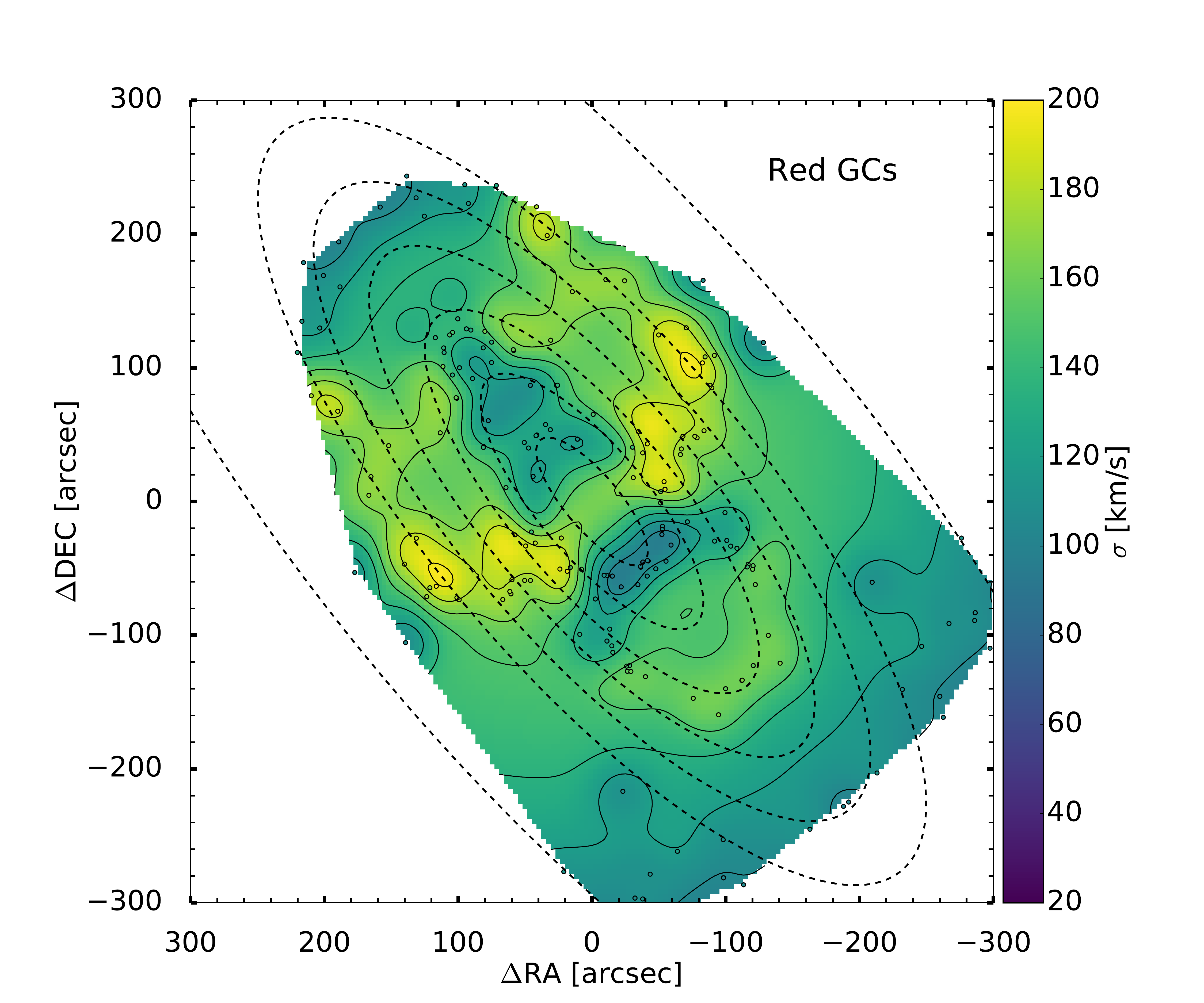}
\end{figure*}
\begin{figure*}
%    \ContinuedFloat
    \centering
    \begin{tabular}{cc}
    \includegraphics[width=0.48\textwidth]{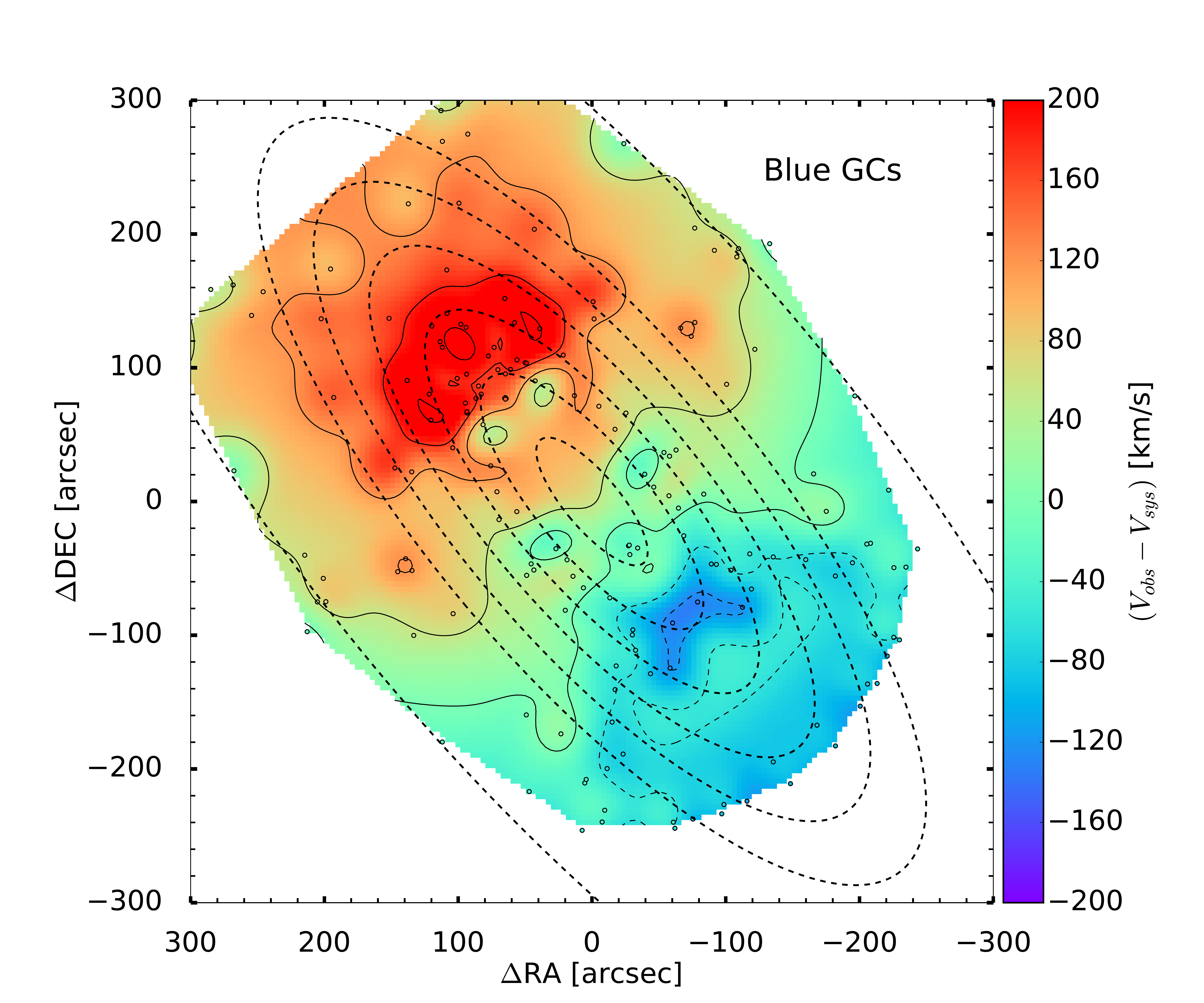}
    \includegraphics[width=0.48\textwidth]{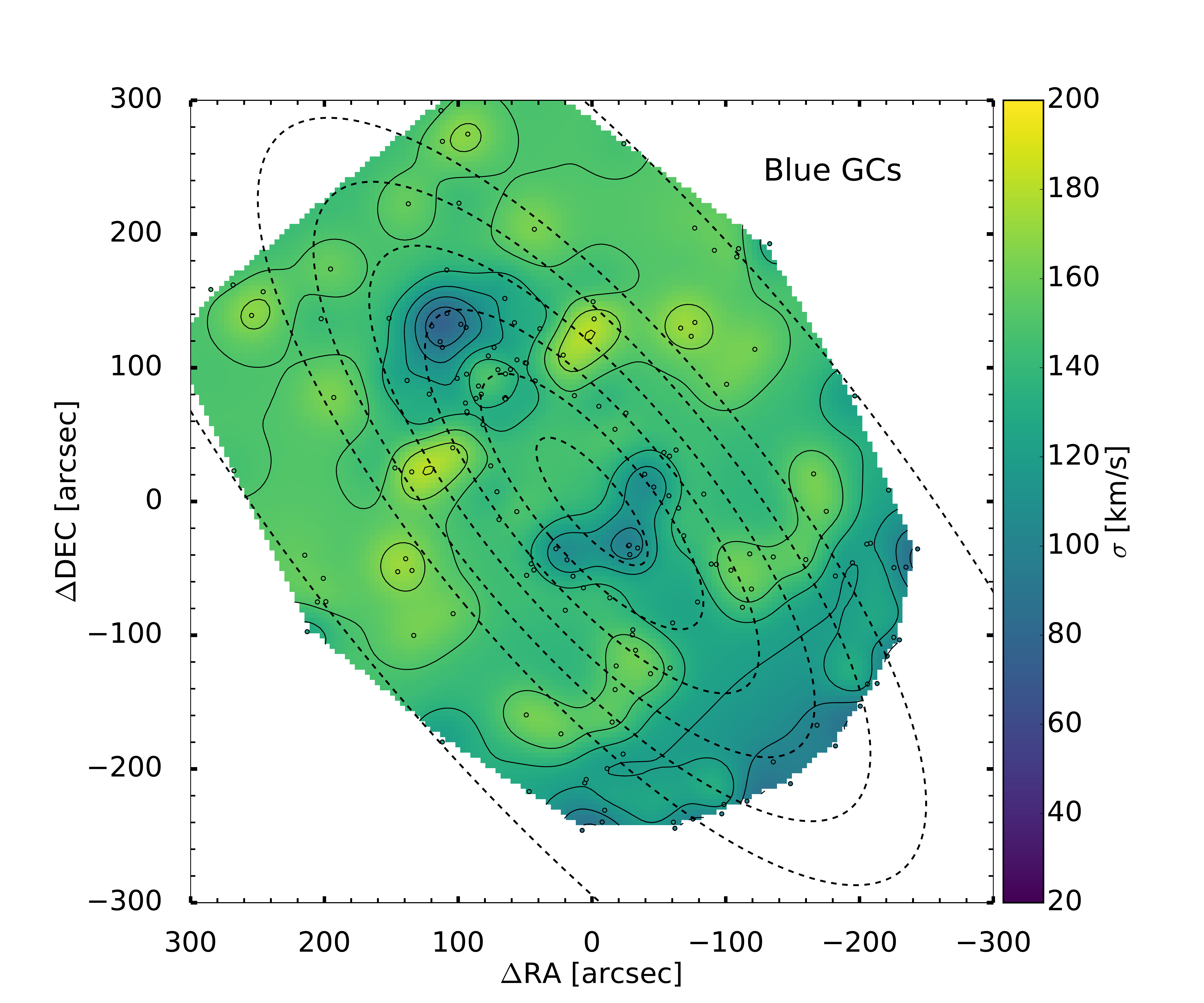}
    \end{tabular}
    \caption{From top to bottom: 2D velocity (left) and velocity dispersion (right) kinematic maps of the SLUGGS stars (integrated starlight), PNe, red and blue GCs of NGC 3115. The kriging maps are produced using the \textit{folded} catalogs with respect to the photometric major-axis of the galaxy. The small open circles indicate the positions of the tracers where their kinematics were extracted and the black solid lines represent the iso-velocity contours. The black dashed ellipses are the $1-6\, R_\mathrm{e}$ and $10\, R_\mathrm{e}$ photometric ellipses (Table \ref{tab:NGC3115_parameters}). The 2D maps are normalised with respect to the velocity scale bar on the right.}
    \label{fig:2D_KrigingMaps}
\end{figure*}

\subsubsection{The GCs rotation field vs. UCD velocities}
\label{sec:GCvsUCDs_2D}
We now compare the velocities of the UCDs of NGC 3115 with the velocity field of the full (red $+$ blue) GC catalogue.
We include the UCDs from our KCWI spectral analysis ($7$ objects including UCD 17a) and the $5$ from the literature \citep{Pota2013,Jennings2014} for a total of $12$ objects.

\begin{figure}
    \centering
    \includegraphics[width=0.5\textwidth]{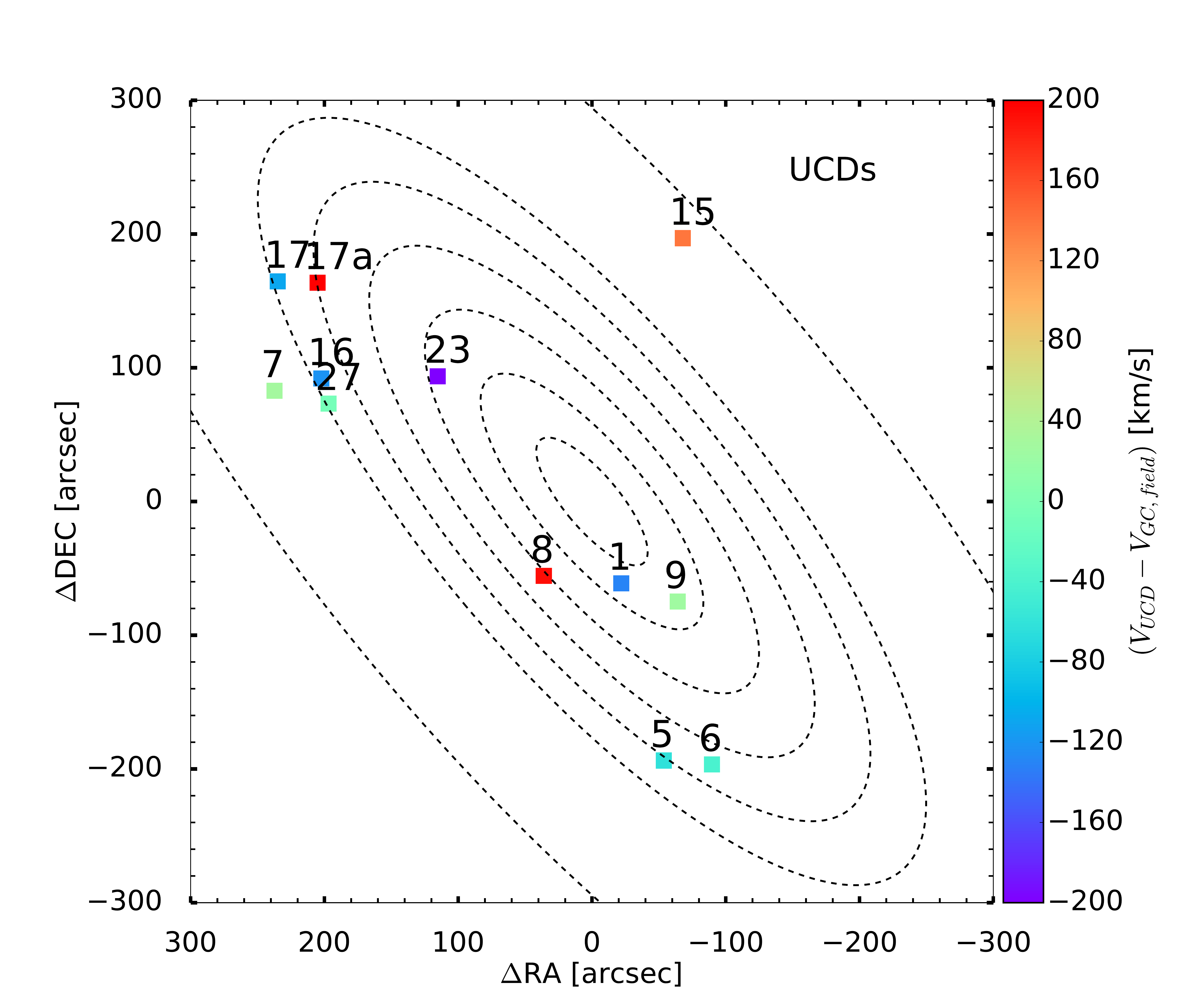}
    \caption{2D map of the UCD velocities relative to the underlying GC velocity field. This 2D map is similar in structure to the 2D maps in Fig. \ref{fig:2D_KrigingMaps}. The UCD velocities are normalized by $(V_{\mathrm{UCD}} - V_{\mathrm{GC, field}})$, where $V_{\mathrm{GC, field}}$ is the \textit{folded} red$+$blue GC velocity field. We remind the reader that the figure is oriented North towards the positive $\Delta \mathrm{DEC}$ direction and East towards the positive $\Delta \mathrm{RA}$ direction. This means that the underlying GC velocity field has positive velocity, i.e. receding with $(V_{\mathrm{obs}}-V_{\mathrm{sys}})>0\, \mathrm{km^{-1}}$, toward North-East, while it has negative velocity, i.e. approaching with $(V_{\mathrm{obs}}-V_{\mathrm{sys}})<0\, \mathrm{kms^{-1}}$, toward South-West.}
    \label{fig:AllGCsplusUCDs_1D}
\end{figure}

\begin{figure*}
    \centering
    \includegraphics[width=0.85\textwidth]{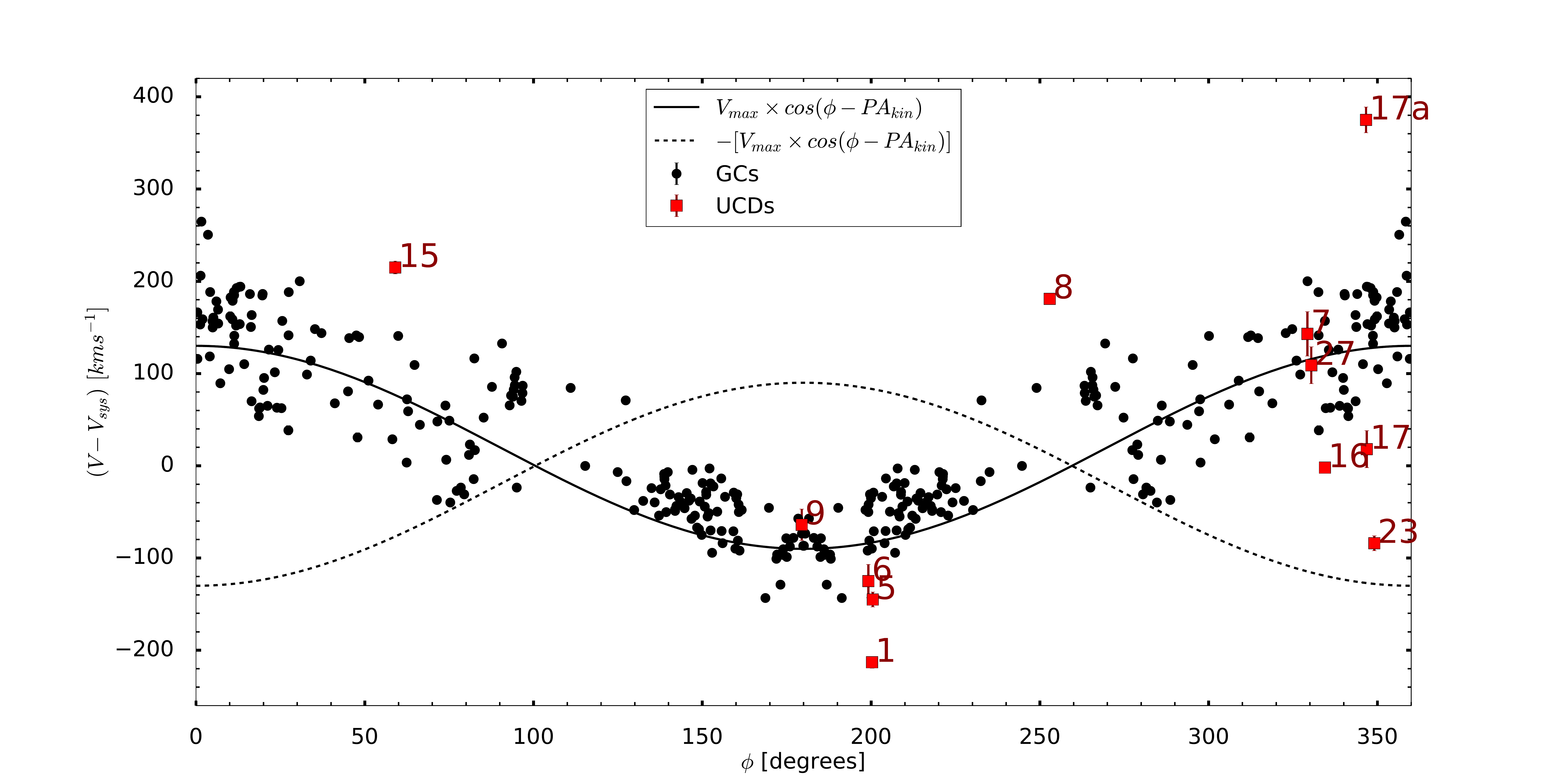}
    \caption{Red$+$blue GCs (black points) and UCDs (red points) radial velocities ($V_{\mathrm{obs}} - V_{\mathrm{sys}}$ as a function of the position angle, $\Phi$, of the objects. As black solid line, we show the rotation model curve, given by Eq. \ref{eq:cosine_law3}, fitted to all the GCs. The dashed black line represents the corresponding counter-rotating model curve. The GCs catalogue used here corresponds to the \textit{folded} one which we used to produce the 2D velocity maps shown in Fig. \ref{fig:2D_KrigingMaps}. We highlight UCD 23, which is counter-rotating with respect to the underlying GC velocity field consistently with the counter-rotating model rotation curve.}
    \label{fig:AllGCsplusUCDs_1D_Original}
\end{figure*}

In Fig. \ref{fig:AllGCsplusUCDs_1D}, we show the 2D map of the UCD velocities relative to the underlying GC velocity field, i.e. the UCD velocities are normalized by $V_{\mathrm{UCD}} - V_{\mathrm{GC, field}}$. Here, the observed velocity of each UCD, $V_{\mathrm{obs}}$, is already subtracted by the $V_{sys}$ of NGC 3115 (see Table \ref{tab:NGC3115_parameters}). A $(V_{\mathrm{UCD}}-V_{\mathrm{GC, field}}) \sim 0\, \mathrm{kms^{-1}}$ means that the UCD velocity is consistent with the underlying GC velocity field at the UCD location and therefore the UCD is following the rotation of the underlying GCs.
The larger the absolute difference $(V_{\mathrm{UCD}}-V_{\mathrm{GC, field}})$ is, the less the UCD velocity is consistent with the underlying GC velocity field, meaning that the UCD is oddly rotating with respect to the underlying GCs.
In a similar way, the corresponding 1D plot in Fig. \ref{fig:AllGCsplusUCDs_1D_Original} quantifies the difference between the UCD and GC velocities. Fig. \ref{fig:AllGCsplusUCDs_1D_Original} shows the red $+$ blue GCs (black points) and the UCDs (red squares) radial velocities as a function of the position angle, $\Phi$, of the objects. This figure is comparable in structure to Fig. \ref{fig:point_symmetry_catalogs}. The black solid line represents the rotation model curve, given by the simple cosine rotation law of Eq. \ref{eq:cosine_law3}, fitted to the combined GC velocities with the corresponding counter-rotating model curve shown as dashed black line.

From Fig. \ref{fig:AllGCsplusUCDs_1D} and \ref{fig:AllGCsplusUCDs_1D_Original}, we see that UCD 5, 6, 7, 9, 16, 17 and 27 are consistent with the underlying GC velocity field, as they all lie within the GC scatter from the best-fit model rotation curve (see Fig. \ref{fig:AllGCsplusUCDs_1D_Original}), while the other UCDs are not.
In Fig. \ref{fig:AllGCsplusUCDs_1D_Original}, we highlight one object, UCD 23, which is counter-rotating with $(V_\mathrm{obs}-V_\mathrm{sys})=-84\, \mathrm{km\,s^{-1}}$ with respect to the underlying GCs. 
UCD 15, between $8-10\, R_\mathrm{e}$, is instead much faster rotating with $(V_\mathrm{obs}-V_\mathrm{sys})=215\, \mathrm{km\,s^{-1}}$ than the underlying GCs in that region, i.e. $\sim60\, \mathrm{km\,s^{-1}}$.
Similarly, UCD 17a shows a much higher velocity, $(V_\mathrm{obs}-V_\mathrm{sys})=375\, \mathrm{km\,s^{-1}}$, than that of the GCs (i.e. $\sim120\, \mathrm{km\,s^{-1}}$). UCD 1, close to $\sim1\, R_\mathrm{e}$, is also faster approaching towards us with $(V_\mathrm{obs}-V_\mathrm{sys})=-213\, \mathrm{km\,s^{-1}}$ than the underlying GCs at $\sim-100\, \mathrm{km\,s^{-1}}$.
Finally UCD 8, at $\sim3\, R_\mathrm{e}$ along the minor-axis, displays a significant rotation with $(V_\mathrm{obs}-V_\mathrm{sys})=181\, \mathrm{km\,s^{-1}}$ relative to the underlying GCs, which do not show a net rotation along the minor-axis.

% -------------------- The Model ---------------------- %
\subsection{Kinemetry}
\label{sec:kinemetry}
In order to model our kinematic data and to facilitate comparison between the integrated kinematics of the stars with the discrete kinematics of GCs and PNe, we use the method developed by \citet{Krajnovic2006} called \texttt{kinemetry}\footnote{\url{http://davor.krajnovic.org/idl/}}.

We have seen in Fig. \ref{fig:point_symmetry_catalogs} that all our kinematic catalogs are well described by the simple cosine law given in Eq. \ref{eq:cosine_law3} and that they are axi-symmetric with respect to the major-axis of the galaxy. From Sec. \ref{sec:2D_kinematic_maps}, we have seen that the overall $\mathrm{PA}_\mathrm{kin}$ is also well aligned with the $\mathrm{PA}_\mathrm{phot}$ of NGC 3115, which is typically expected for fast-rotating early-type galaxies and for axisymmetric systems in dynamical equilibrium \citep{Statler1994a,ATLAS3D2011,Krajnovic2011}.
So, to apply \texttt{kinemetry} to fit our kinematic tracers, which include PNe and GCs, as well as the stellar kinematic data described in Secs. \ref{sec:stellar_kinematic}--\ref{sec:planetary_nebulae}, we adopt a similar procedure as in \citet{Proctor2009,Foster2016,Bellstedt2017} for fitting sparse and non-homogeneous datasets.
We manually bin our data in elliptical annuli of varying sizes such that each bin contains $10$ data points. Then we extract the best-fitting kinematic moments at the radius of each bin by fitting Eq. \ref{eq:cosine_law} and Eq.~\ref{eq:cosine_law2} within \texttt{kinemetry}. In this way, we compromise between having enough data-points in each bin and a sufficient number of bins to reliably extract the kinematic moments at the different radii. For our purposes, we focus only on recovering the first and second order moments, $V_\mathrm{rot}$ and $\sigma$, of the line-of-sight velocity distribution (LOSVD) as, due to the sparse nature of our kinematic catalogs, we do not have enough signal-to noise to accurately calculate the higher order moments, $h_{n}$ ($n>2$), of the LOSVD as well.
The \texttt{kinemetry} chi-squared minimization process to calculate the best-fitting kinematic moments is carried out separately 
for each radial ellipse, starting from the central annulus and moving outward until the edges of the map are reached, and is explained in detail in \citet{Foster2016}.

When running \texttt{kinemetry} with sparse and non-homogeneous datasets, it is common to fix the value of the kinematic axial-ratio to the photometric one, i.e. $q_\mathrm{kin}=q_\mathrm{phot}$ (e.g. \citealt{Arnold2011,Foster2013,Foster2016,Bellstedt2017}), as this provides better stability in the results. In fact, $q_\mathrm{kin}$ is not well constrained in the case of sparse datasets, even when the data are well-sampled and of good quality. 
In their work, \citet{Foster2016} mentioned that the choice of fixing $q_\mathrm{kin}$ to the photometric value can underestimate the recovered $V_\mathrm{rot}$ profile by up to $\sim10$--$20\%$. 

In this work, we first run \textit{kinemetry} on the 2D MUSE stellar velocity and velocity dispersion maps of NGC 3115 from \citet{Guerou2016}. These MUSE data are spatially resolved continuous maps of kinematic measurements around the galaxy out to $\sim2.5\, R_\mathrm{e}$. These well-sampled data allow us to calculate also the best-fitting kinematic parameters, $\mathrm{PA}_\mathrm{kin}$ and $q_\mathrm{kin}$, at the different radii together with the rotation velocity and velocity dispersion moments. 
In doing so, we find that the $\mathrm{PA}_\mathrm{kin}$ is very well aligned with the $\mathrm{PA}_\mathrm{phot}$ of NGC 3115 (see Table \ref{tab:NGC3115_parameters}) at all radii probed by \citet{Guerou2016}. We also find that $q_\mathrm{kin}\sim0.2$ 
at $\sim0.4\, R_\mathrm{e}$, while $q_\mathrm{kin}\sim0.45$ beyond $\sim0.5\, R_\mathrm{e}$ and out to $\sim2.5\, R_\mathrm{e}$ where the rotation velocity flattens to constant values. 
The velocity dispersion has the opposite trend in the sense that it has a maximum in the central regions within $\sim0.2\, R_\mathrm{e}$ and then it flattens to lower values out to $\sim2.5\, R_\mathrm{e}$.

When running \textit{kinemetry} on the SLUGGS stars, we find that both the $q_\mathrm{kin}$ and $\mathrm{PA}_\mathrm{kin}$ are still well constrained to $\sim0.5$ and $\sim45\degr$, respectively. The $q_\mathrm{kin}$ value also agrees well with the one we recover from the MUSE \citep{Guerou2016} maps between $\sim1-2.5\, R_\mathrm{e}$, i.e. the overlapping radii of these two datasets, even though it has much larger errors indicative of the fact that it is not as well constrained as from the MUSE data. The $\mathrm{PA}_\mathrm{kin}$ value is, instead, well constrained and it suggests a small misalignment with respect to the \citet{Guerou2016} stellar data and $\mathrm{PA}_\mathrm{phot}$ of NGC 3115.

On the contrary, $q_\mathrm{kin}$ is less well constrained for the GCs and PNe when left free to vary. 
We perform some tests with \textit{kinemetry} to check the effect of fixing $q_\mathrm{kin}$ on our kinematic profiles. 
We find that $q_\mathrm{kin}$ has no significant impact on the shape of the rotation velocity profile, and only the amplitude varies.
However, we find that the amplitude difference of the rotation velocity is not significant, i.e. $\lesssim5\, \mathrm{km\,s^{-1}}$, and is contained within the $1$-$\sigma$ uncertainties.
For this reason, we fix $q_\mathrm{kin}=0.5$ for calculating the rotation velocity and velocity dispersion profiles of both the red and blue GCs and PNe. This is also the value constrained from the stellar kinematics beyond $\sim0.5\, R_\mathrm{e}$, which provides the most stable and smooth results for the GCs and PNe tracers.

We also test the impact of a fixed $\mathrm{PA}_\mathrm{kin}$ as opposed to leaving it free to vary. We find that both the shape and amplitude of the recovered rotation velocity and velocity dispersion profiles do not vary significantly. 
Specifically, we find that the blue GCs have a good kinematic alignment with the overall $\mathrm{PA}_\mathrm{phot}$ of the galaxy, while the red GCs and PNe display very similar $\mathrm{PA}_\mathrm{kin}$ profiles with misalignment of up to $\sim(20\pm10)\degr$ between $\sim2-4\, R_{\mathrm{e}}$ with respect to the $\mathrm{PA}_\mathrm{phot}$ of the galaxy and the $\mathrm{PA}_\mathrm{kin}$ of the MUSE and SLUGGS stellar kinematics.

\begin{figure*}
    \includegraphics[width=1.05\textwidth]{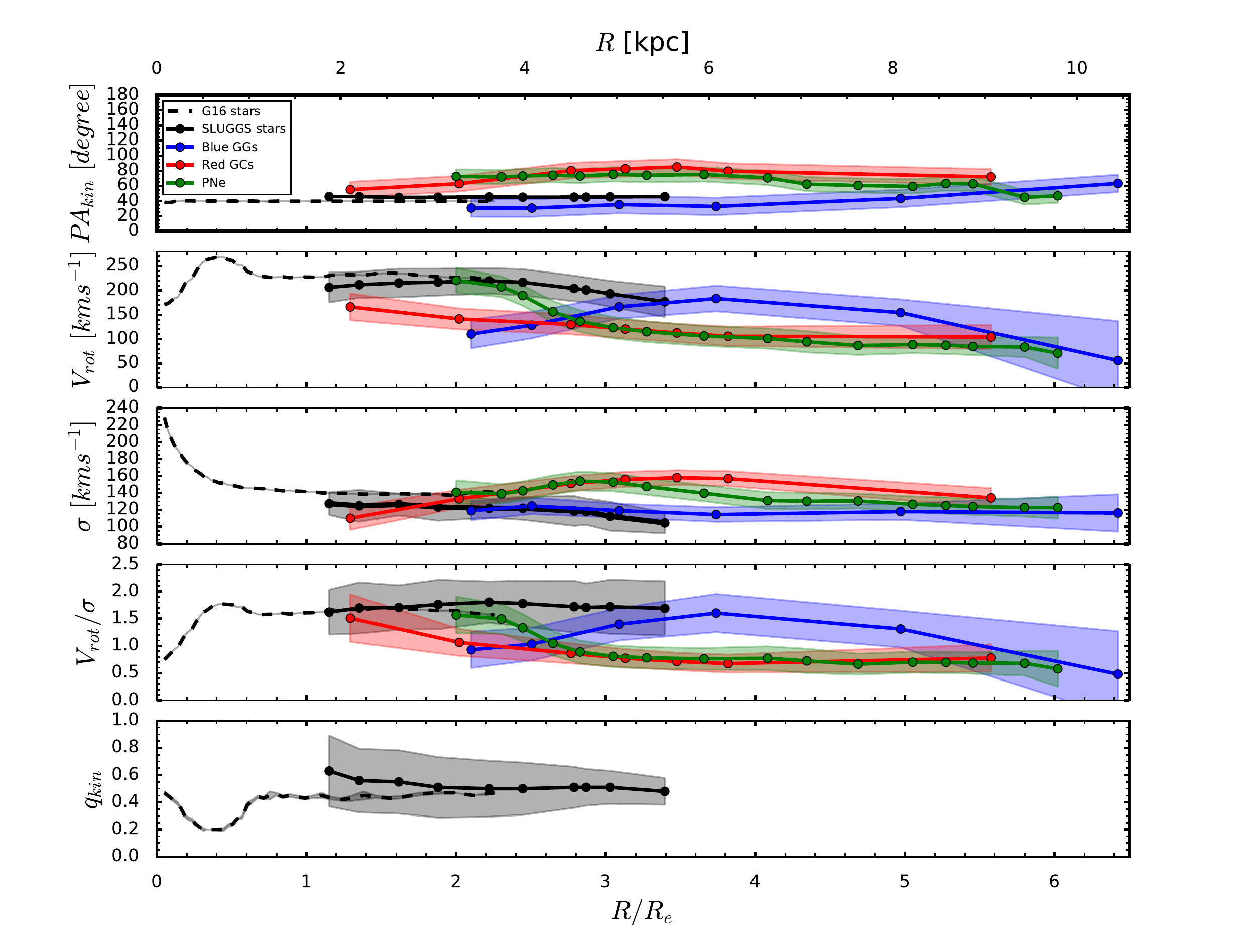}
    \caption{From top to bottom: 1D profiles of the kinematic position angle, $\mathrm{PA}_{\mathrm{kin}}$, the rotation velocity, $V_\mathrm{rot}$, the velocity dispersion, $\sigma$, the rotation velocity to velocity dispersion ratio, $V_\mathrm{rot}/\sigma$ and the kinematic axial-ratio, $q_{\mathrm{kin}}$, for the PNe (green line), blue and red GCs (blue and red lines, respectively), SLUGGS (black line) and MUSE (i.e. G16 for \citealt{Guerou2016}; dashed line) stars (integrated starlight). 
    The $q_{\mathrm{kin}}$ is only shown for the SLUGGS and MUSE data as it is kept fixed to $q_{\mathrm{kin}}=0.5$ for the red/blue GCs and PNe.
    The radial kinematics are recovered from fitting along radial ellipses using the \textit{kinemetry} program. The shaded regions represent the $1$-sigma uncertainties calculated from running $500$ bootstrapped samples obtained by sampling with replacement each one of our kinematic dataset (similarly to \citet{Foster2013}). Overall, the profiles show three smooth kinematic transitions: from the dispersion-dominated ($V_\mathrm{rot}/\sigma<1$) bulge at $\sim0.2\, R_\mathrm{e}$, to a fast-rotating disk and finally to a slowly rotating spheroid ($V_\mathrm{rot}/\sigma<1$) seen at $\sim2$--$2.5\, R_\mathrm{e}$ for the red GCs and PNe, and 
    at $\sim5\, R_\mathrm{e}$ for the blue GCs.}
    \label{fig:kinemetry_results}
\end{figure*}

Fig. \ref{fig:kinemetry_results} shows the results of our  \textit{kinemetry} fits performed on each kinematic dataset.
For the GC and PNe, we use the averaged velocity and velocity dispersion measurements from the nearest-neighbour binning, as described in Sec. \ref{sec:2D_kinematic_maps} for the 2D kinematic maps. 
In this plot, from top to bottom, we show the kinematic position angle, $\mathrm{PA}_{\mathrm{kin}}$, the recovered rotation velocity, $V_\mathrm{rot}$, the velocity dispersion, $\sigma$, the rotation parameter, $V_\mathrm{rot}/\sigma$, and the kinematic axial-ratio, $q_{\mathrm{kin}}$, for the stellar kinematics from SLUGGS and MUSE \citep{Guerou2016}, as it was fixed for the red/blue GCs and PNe, as a function of the ratio $R/R_{e}$. 

We observe good agreement between the kinematic profiles of the MUSE \citep{Guerou2016} and the SLUGGS stellar data at $\sim1$-$2\, R_\mathrm{e}$ where they overlap.
We find evidence of an offset of $\sim20\, \mathrm{km\,s^{-1}}$ between the \citet{Guerou2016} and SLUGGS stellar velocity dispersion profiles in the overlapping region, similarly to the offsets found by \citet{Foster2016} between the ATLAS 3D and SLUGGS stellar kinematic data. This offset was discussed by \citet{Foster2016} to be likely due to the more aggressive binning performed in ATLAS 3D in the outskirts of galaxies to reach the desired target S/N. For the MUSE stellar kinematics, \citet{Guerou2016} did perform a similar spatial binning in order to reach a target S/N $=50$ per $\mathrm{\angstrom}$ in each spatial bin. However, the velocity dispersion offset that we observe in Fig. \ref{fig:kinemetry_results} is included within the $1$-sigma uncertainties of the SLUGGS stars.
The $V_\mathrm{rot}$ profile shows a steep rise to the maximum rotation of $\sim270\, \mathrm{km\,s^{-1}}$ at $\sim0.4\, R_\mathrm{e}$ and then it remains constant to $\sim220\, \mathrm{km\,s^{-1}}$ beyond $\sim0.6\, R_\mathrm{e}$. The $\sigma$ profile has a very high peak at $\sim240\, \mathrm{km\,s^{-1}}$ in the central regions (within $0.2\, R_\mathrm{e}$) after which it decreases and remains constant to $\sim140\, \mathrm{km\,s^{-1}}$ until $\sim2\, R_\mathrm{e}$. 
There is a hint of decrease in the stellar rotation velocity profile beyond $\sim1.8\, R_\mathrm{e}$, which is slightly visible in \citet{Guerou2016} data and continues in the complementary SLUGGS stellar data out to $\sim3.5\, R_\mathrm{e}$. This feature is, however, not visible in the stellar $V_\mathrm{rot}/\sigma$ profile which remains roughly flat out to $\sim3.5\, R_\mathrm{e}$ as both the rotation velocity and velocity dispersion decrease together over the range 2--$3\, R_\mathrm{e}$.

The decrease of the rotation velocity, as well as of $V_\mathrm{rot}/\sigma$, at larger radii is more evident in the GC and PNe tracers, extending to $\sim6.5\, R_\mathrm{e}$. 
We observe that the red GCs and PNe have very similar kinematic profiles beyond $\sim2.5\, R_\mathrm{e}$. There is a small amplitude mismatch in the $V_\mathrm{rot}$ and $V_\mathrm{rot}/\sigma$ profiles of the red GCs and PNe at $\sim2$--$2.5\, R_\mathrm{e}$, with the PNe displaying higher rotation which very well agrees with that of the stars in that region. 

Specifically, the red GCs display a smooth transition from $V_\mathrm{rot}/\sigma>1$ to $V_\mathrm{rot}/\sigma<1$ at $\sim2\, R_\mathrm{e}$, with $V_\mathrm{rot}$ reaching its maximum ($\sim170\, \mathrm{km\,s^{-1}}$) between $\sim1-2\, R_\mathrm{e}$. The $\sigma$ has the opposite trend with higher dispersion beyond $\sim2\, R_\mathrm{e}$, where the $V_\mathrm{rot}$ goes down.
The PNe also display the same smooth transition of the $V_\mathrm{rot}/\sigma$, as the red GCs, occurring at $\sim2.5\, R_\mathrm{e}$, beyond which both PNe and red GCs have overall flat $V_\mathrm{rot}/\sigma\sim0.5$ profiles out to $\sim6\, R_\mathrm{e}$. The velocity dispersion of the PNe has its peak at $\sim3\, R_\mathrm{e}$, which roughly matches with the transition region from $V_\mathrm{rot}/\sigma>1$ to $V_\mathrm{rot}/\sigma<1$, and its profile is also overall consistent with that of the red GCs.   

On the other hand, the blue GCs show a slightly different rotation velocity profile with higher amplitude at $\sim3$--$5\, R_\mathrm{e}$. The velocity dispersion remains overall flat at all radii out to $\sim6.5\, R_\mathrm{e}$. As a result, the corresponding $V_\mathrm{rot}/\sigma$ profile is very similar to the $V_\mathrm{rot}$ profile. 
Although the uncertainties are greater at large radii, there is an indication that the blue GCs also transition from $V_\mathrm{rot}/\sigma<1$ to $V_\mathrm{rot}/\sigma>1$ at $\sim3\, R_\mathrm{e}$, which then transitions back to $V_\mathrm{rot}/\sigma<1$ beyond $\sim5\, R_\mathrm{e}$ and out to $\sim6.5\, R_\mathrm{e}$, following a similar behaviour as the red GCs and PNe out to $\sim6\, R_\mathrm{e}$.

We emphasize that our 1D kinematic results are overall in good agreement with the previous 2D velocity and velocity dispersion maps (see Sec. \ref{sec:2D_kinematic_maps}).

% -------------------- Discussions and Conclusions ----------------- %
\section{Discussion}
\label{sec:Discussions}

\subsection{NGC 3115 kinematics}
\label{sec:N3115_kinematics}
In Sec. \ref{sec:results}, we studied the kinematic properties of NGC 3115 out to $\sim6.5\, R_\mathrm{e}$ from the 2D velocity and velocity dispersion maps and 1D profiles of our discrete tracers (red, blue GCs and PNe) and stars.

% First Point
We find that all our discrete tracers, as well as the stars, are consistent with being axi-symmetric (see Sec. \ref{sec:PointSymmetry}) and show a strong rotation which is overall aligned with the $\mathrm{PA}_\mathrm{phot}$ of NGC 3115 (see Table \ref{tab:NGC3115_parameters}) with small kinematic misalignment of maximum $\sim5\degr$ for the SLUGGS stars and {\bf $\sim10-20\degr$} for the red GCs and PNe (see Sec. \ref{sec:kinemetry}). 
A similarly small kinematic misalignment was found for the PNe tracers by \citet{Pulsoni2018Survey}. 
These results suggest that NGC 3115 has reached a dynamical equilibrium state with its discrete tracers, i.e. GC sub-populations and PNe, which are co-rotating with the stars with similar velocity amplitudes as shown in Fig. \ref{fig:2D_KrigingMaps} and Fig. \ref{fig:kinemetry_results}. 

% Second Point
We find good overall agreement between the red GC and the PNe kinematic profiles. The stellar (\citet{Guerou2016} $+$ SLUGGS) $V_\mathrm{rot}$ profiles have similar amplitude to the PNe at $\sim2\, R_\mathrm{e}$, beyond which they decline out to $\sim3.5\, R_\mathrm{e}$ following the rotation velocity profile of both the red GCs and PNe, even though the $V_\mathrm{rot}/\sigma$ remains flat (see Fig. \ref{fig:kinemetry_results}).
This suggests that both the red GCs and PNe are tracing the kinematics of the underlying stellar population, which is in agreement with previous observational works that found that both tracers follow the stellar surface brightness profile \citep{Coccato2009,Cortesi2013a,Zanatta2018} and therefore, for this reason, they can be effectively used as proxies for the stars in galaxy kinematic studies at large radii \citep{Romanowsky2006,Hartke2017,Forbes2018}.

At $\sim2-2.5\, R_\mathrm{e}$ we identify a smooth transition from disk-like ($V_\mathrm{rot}/\sigma>1$) to spheroid-like ($V_\mathrm{rot}/\sigma<1$) kinematics for both the red GCs and PNe. The red GC and PNe population located beyond $\sim2-2.5\, R_\mathrm{e}$ are likely dominated by the slower rotating ($V_\mathrm{rot}/\sigma<1$) spheroidal component of NGC 3115, as found by \citet{Cortesi2013b} for the PNe.
At smaller radii, (within $\sim2$--$2.5\, R_{\mathrm{e}}$) red GCs and PNe are likely dominated by the stellar disk component, given that their kinematics is very similar to that of the stars.
The first transition from spheroid-like to disk-like kinematics occurs at $\sim0.2\, R_\mathrm{e}$ and is visible from the \citet{Guerou2016} stellar kinematics (see Fig. \ref{fig:kinemetry_results}).

The blue GCs display different $V_\mathrm{rot}$ and $V_\mathrm{rot}/\sigma$ profiles (the velocity dispersion remains roughly flat at all radii) compared to the red GCs, having higher amplitudes at larger radii ($\sim3$--$5\, R_\mathrm{e}$; see Fig. \ref{fig:kinemetry_results}). We also see this in the corresponding 2D maps in Fig. \ref{fig:2D_KrigingMaps}. Previous work on the GC kinematics of NGC 3115  found this difference between the 1D $V_{\mathrm{rot}}/\sigma$ profiles of the red and blue GC sub-populations (e.g. \citealt{Arnold2011,Pota2013,Zanatta2018}).
\citet{Zanatta2018} 
found $V_{\mathrm{rot}}$ profiles for the red, blue GCs and PNe that differ slightly from our results. 
They found that the red GCs and PNe have similar profiles out to large radii, but the $V_{\mathrm{rot}}$ amplitude of the PNe is slightly higher and more similar to that of the blue GCs than the red GCs.
The difference between \citet{Zanatta2018} and our results could be due to the use of slightly different GC catalogues and to their bulge-disk  likelihood analysis in which they rejected some GC objects. 
However, the results are consistent when comparing the $V_{\mathrm{rot}}/\sigma$ profiles of the red and blue GC sub-populations. For the PNe, \citet{Zanatta2018} did not show the $V_{\mathrm{rot}}/\sigma$ profile. 

The different kinematic behaviour 
between the blue and the red GC sub-populations 
could suggest a physical difference. 
As blue GC systems are typically more extended than the red ones (i.e. the blue GCs are located at larger physical distances from the galaxy), projection effects would shift velocities towards larger radii. Thus projection effects would tend to increase the observed difference between the blue and the red GCs kinematics profiles. 

Additionally, while the red GCs are generally expected to have formed \textit{in-situ} within the host galaxy and so they should trace the kinematic properties of the underlying stellar population, the blue GCs are generally expected to have formed \textit{ex-situ} and accreted onto the host galaxy during mergers with satellite dwarf galaxies and so they should trace the properties of the galaxy halo (e.g. \citealt{Remus2018,Forbes2018}).
However, this does not exclude that a fraction of the blue GCs also formed \textit{in-situ} within the galaxy. 
Indeed \citet{Choksi2019} predicted for similar stellar masses as NGC 3115 (Table \ref{tab:NGC3115_parameters}), that around half of the blue GCs are accreted, while the rest have an \textit{in-situ} origin within the galaxy (see their figure 6). The red GCs are, in contrast, expected to have almost all formed \textit{in-situ} with just a small fraction accreted. 

The disk-like kinematics ($V_\mathrm{rot}/\sigma>1$) displayed by the blue GCs between $\sim3-5\, R_\mathrm{e}$ and the subsequent possible smooth transition to spheroid-like kinematics ($V_\mathrm{rot}/\sigma<1$) beyond $\sim5\, R_\mathrm{e}$ (see Fig. \ref{fig:kinemetry_results}) suggests that the blue GC sub-population within $\sim3-5\, R_\mathrm{e}$ could be dominated by the \textit{in-situ} formed component of NGC 3115, which is tracing the stellar population of the fast-rotating disk. The blue GC sub-population beyond $\sim5\, R_\mathrm{e}$ would, instead, be tracing the outer slowly rotating spheroidal component, similarly to the red GC and PNe tracers in that region, with a fraction of this blue population likely accreted (i.e. \textit{ex-situ} origin). 

In Table \ref{tab:summary_transitions}, we summarise the radius at which the main changes from disk-like ($V_\mathrm{rot}/\sigma>1$) to spheroid-like ($V_\mathrm{rot}/\sigma<1$) kinematics occur in our discrete tracers and stellar profiles in Fig. \ref{fig:kinemetry_results}. 

\begin{table}
\caption{Radii at which we identify a change in the $V_\mathrm{rot}/\sigma$ profiles of our discrete tracers and stars in Fig. \ref{fig:kinemetry_results}. \textit{Column 1:} kinematic datasets, \textit{column 2:} positive gradient of $V_\mathrm{rot}/\sigma$, \textit{column 3:} negative gradient of $V_\mathrm{rot}/\sigma$. A positive gradient means a change from spheroid-like ($V_\mathrm{rot}/\sigma<1$) to disk-like ($V_\mathrm{rot}/\sigma>1$) kinematics and vice-versa for a negative gradient.}
\centering
\begin{tabular}{|l|c|c|}\hline\hline
Data     & $\Delta V_\mathrm{rot}/\sigma>0$ & $\Delta V_\mathrm{rot}/\sigma<0$    \\ \hline
Stars    & $\sim0.2\, R_\mathrm{e}$  &                            \\
Red GCs  &                             & $\sim2\, R_\mathrm{e}$   \\ 
PNe      &                             & $\sim2.5\, R_\mathrm{e}$ \\
Blue GCs &                             & $\sim5\, R_\mathrm{e}$   \\
\hline \hline
\end{tabular}
\label{tab:summary_transitions}
\end{table}

Overall, our kinematic results of Fig. \ref{fig:kinemetry_results} suggest the presence of a fast-rotating disk embedded in a slower rotating spheroidal component in NGC 3115 (e.g. \citealt{Arnold2011,Foster2016,Guerou2016,Zanatta2018,Poci2019}). 

\subsection{Comparison to simulations of galaxy formation}
\label{sec:Simulations}
From hydro-dynamical simulations of early-type galaxies with stellar masses of $2\times10^{10}\, \mathrm{M_{\odot}} < M_{*} < 6\times10^{11}\, \mathrm{M_{\odot}}$, \citet{Naab2014} identified six different classes of galaxies from their present-day shapes and kinematic properties that formed through different evolutionary paths. According to their results, fast rotator galaxies may be a result of either gas-rich minor or major mergers or late gas-poor major mergers, i.e. \textit{class A, B, D} in \citet{Naab2014}, respectively. However only the remnants of gas-rich mergers show evidence of an embedded cold disk feature along the major-axis, as we observe in NGC 3115, and this feature is stronger in gas-rich minor merger (mass-ratio $<$ 1:4) remnants (\textit{class A}) than in gas-rich major merger (mass-ratio $>$ 1:4) ones (\textit{class B}). 

The observed stellar kinematic properties of NGC 3115 from Fig. \ref{fig:2D_KrigingMaps} and Fig. \ref{fig:kinemetry_results} are consistent with a formation history characterized by gas-rich minor mergers ($<$ 1:4), similarly to what was found also in \citet{Forbes2016a}. These \textit{class A} remnants are classified as regular fast rotators with symmetric velocity fields and display a peaked profile in the stellar spin parameter, $\lambda_{R}$, which is typically used as a proxy for the angular momentum of galaxies and is analogous to the $V_\mathrm{rot}/\sigma$ parameter in this work (see Fig. \ref{fig:kinemetry_results}). 
The simulated velocity dispersion maps of the \textit{class A} remnants show a depression along the major-axis of the galaxy corresponding to the regions of the fast-rotating cold kinematic disk feature, similar to the observed velocity dispersion map of NGC 3115 (see Fig. \ref{fig:2D_KrigingMaps}). 
Additionally, the \textit{class A} merger remnants also show a strong $V_\mathrm{rot}/\sigma$ vs. $h_{3}$ anti-correlation. \citet{Forbes2016a} studied the higher order moments of the LOSVD, i.e. up to $h_{4}$, for the stellar component of a sample of $24$ early-type galaxies and found evidence of an anti-correlation in the $V_\mathrm{rot}/\sigma$ vs. $h_{3}$ in NGC 3115 (see their figure A4), in agreement with an assembly history characterized by gas-rich minor mergers.

Similar conclusions were also reached by \citet{Schulze2020} in their hydro-dynamic cosmological simulations of the stellar kinematics in galaxies out to $\sim5\, R_\mathrm{e}$. Specifically, they find that galaxies with a peak and subsequent decreasing $V_\mathrm{rot}/\sigma$ profile are typically characterized by the accretion of low-mass dwarf galaxies at later epochs (i.e. $z<2$) through minor (1:10 $<$ mass-ratio $<$ 1:4) or mini (mass-ratio $<$ 1:10) mergers that build up the stellar halo and increase the random motion at large radii without altering the already stabilized central rotation of the galaxy, which formed at early epochs (i.e. $z>2$). 

Similarly \citet{Wu2014} studied the stellar LOSVD of a sample of $42$ zoom-in cosmological hydro-dynamical simulations of early-type galaxies out to large radii, i.e. $5\, R_\mathrm{e}$. From their simulated 2D stellar kinematic maps and 1D profiles, they found that gas-rich major and minor mergers can produce the present-day central fast rotator galaxies with rising or flat or slightly decreasing $V_\mathrm{rot}/\sigma$ profiles.
Specifically, gas-rich major mergers are likely to produce fast-rotating systems with pronounced disk component at all radii, while gas-rich minor mergers are more likely to shape the embedded cold disk kinematic feature of galaxies with declining or flat stellar velocity profiles at large radii. On the contrary, a galaxy showing little or no rotation about both its major and minor axes is likely the result of multiple minor mergers. However, \citet{Wu2014} did not mention the mass-ratio range of their minor and major mergers. 

\citet{Bournaud2005} investigated the morphological and kinematic properties of the galaxy remnants resulting from a wide range of mass ratios, i.e. 1:1 to 1:10, and found that mergers with mass ratios between 1:4.5 and 1:10 are able to produce disky remnants, similar to S0s, with hotter disk kinematics, i.e. mean $V_\mathrm{rot}/\sigma\leq1$ for 1:4.5 mass ratios, as well as disky remnants more rotationally supported, i.e. mean $1<V_\mathrm{rot}/\sigma \leq2$ for 1:7 and 1:10 mass ratios, where the 1:10 mass ratios can even produce rotation dominated systems with mean $V_\mathrm{rot}/\sigma \geq2$, referred to as disturbed spirals, defining the transition mass ratio between intermediate and minor mergers. On the contrary, major mergers with mass-ratio 1:1 to 1:3 are rarer and likely to produce elliptical remnants.
That minor mergers may be a possible mechanism for the formation of S0s was also previously studied by \citet{Bekki1998}, even though his simulation was not able to reproduce S0s with the observed thin stellar disks (i.e. embedded disks).

From Fig. \ref{fig:kinemetry_results}, we see stellar $V_\mathrm{rot}/\sigma\sim1-1.8$ at all radii out to $\sim3.5\, R_\mathrm{e}$, except in the central regions within $0.2\, R_\mathrm{e}$ where the system is dispersion-dominated (i.e. $V_\mathrm{rot}/\sigma<1$). The red GCs, PNe and blue GCs have $1 < V_\mathrm{rot}/\sigma < 1.5$ at $\sim1-2\, R_\mathrm{e}$, $2-2.5\, R_\mathrm{e}$ and $\sim3-5\, R_\mathrm{e}$, respectively, with a decrease to $V_\mathrm{rot}/\sigma<1$ beyond $\sim2-2.5\, R_\mathrm{e}$ for the red GCs and PNe and, possibly, beyond $\sim5\, R_\mathrm{e}$ for the blue GCs.

Simulations of GC kinematics 
in galaxy remnants formed through major or minor merger events  
were carried out by \citet{Bekki2005}.
According to their results, major mergers (mass-ratio $\sim$ 1:1 in their simulations) can produce flattened spatial distributions of both the red and blue GC sub-populations, whose major-axes are aligned with that of the stars, and show a significant degree of rotation. Specifically, the rotation is greater in the outer regions, while the velocity dispersion profile is roughly flat or declining with radius. Globally  $V_\mathrm{rot}/\sigma<1$ for both the red and blue GC sub-populations. These latter properties are inconsistent with those that we observe (see Fig. \ref{fig:kinemetry_results}), i.e 
the declining $V_\mathrm{rot}/\sigma$ profile at larger radii, and the different $V_\mathrm{rot}/\sigma$ profiles of the two GC sub-populations.
This argues against a major merger event for the formation of NGC 3115. 

A minor merger (mass-ratio $\sim$ 1:10 in \citealt{Bekki2005} simulations) is however a more plausible scenario as it is predicted to produce a higher rotation velocity ($V_\mathrm{rot}$) in the blue GCs than in the red ones and, at the same time, a small amount of rotation in both GC sub-populations in the outer regions. We do observe a larger $V_\mathrm{rot}$ amplitude of the blue GCs between $\sim3-5\, R_\mathrm{e}$ as compared to the red ones and the decline of the rotation velocity at larger radii in both GC sub-populations (see Fig. \ref{fig:kinemetry_results}).

We note that the simulations of \citet{Bekki2005} are not an exhaustive study,   
and for this reason a more thorough investigation is needed in order to determine how factors such as  mass ratio, orbital eccentricity, relative inclinations, and initial conditions of the GCs effect the final red and blue GC kinematics in the remnant galaxy.
With this caveat in mind, the GC kinematics also appear consistent with an early gas-rich minor merger (mass-ratio $\sim$ 1:4--1:10), as suggested from  simulations of  stellar kinematics \citep{Bournaud2005,Naab2014,Schulze2020}. 

\citet{Schulze2020} concluded that galaxies with embedded rotating disks (which are characterized by a declining $V_\mathrm{rot}/\sigma$ profile at large radius) formed their disks at early times with more recent minor and mini mergers. They do not tend to experience major mergers.

\subsection{NGC 3115 formation scenario}
\label{sec:NGC3115_formation}

\begin{figure*}
    \includegraphics[width=0.8\textwidth]{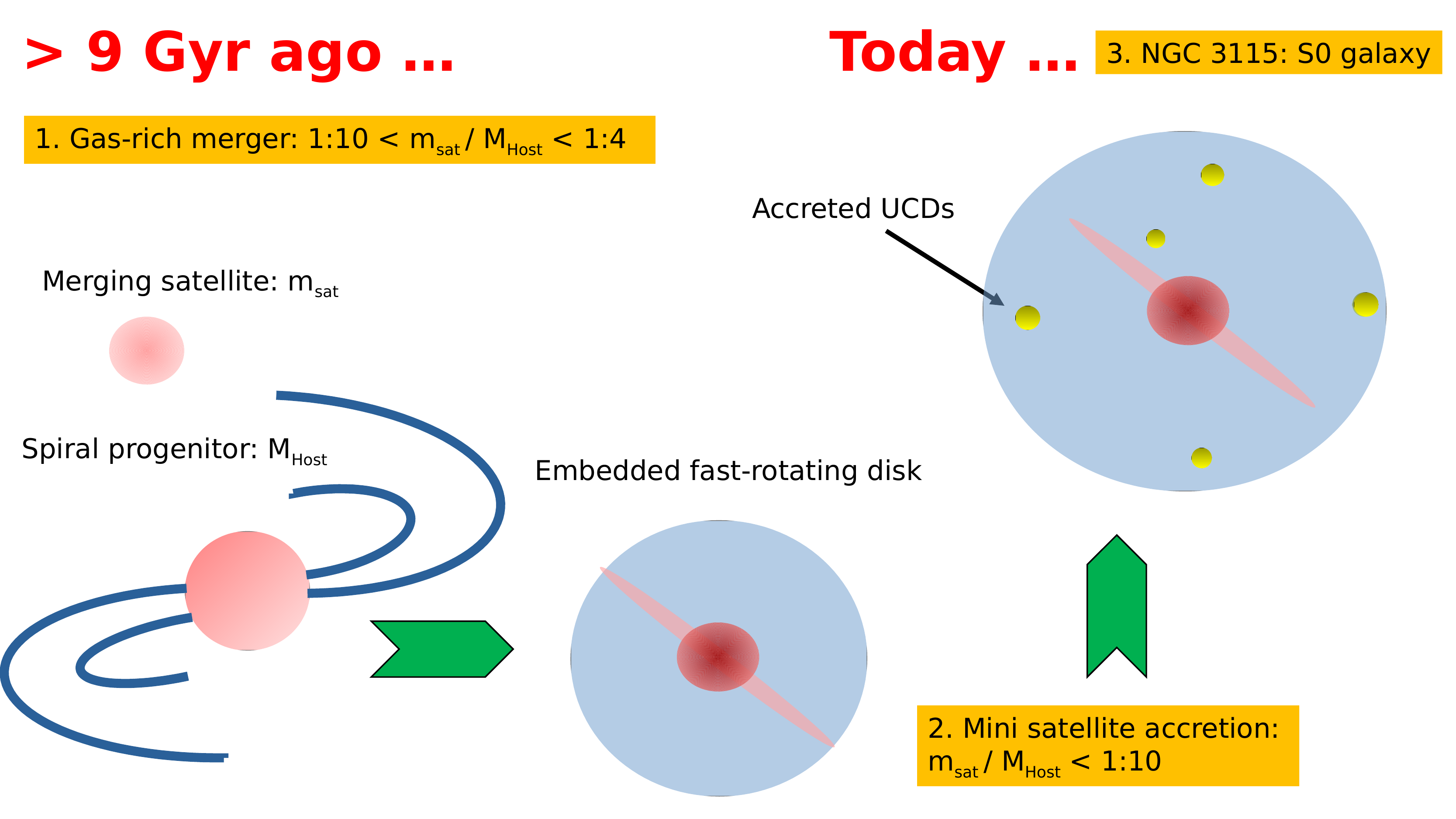}
    \caption{Schematic diagram summarising the main stages that contributed to the formation of  NGC 3115. From left to right, a gas-rich satellite undergoes a minor merger with a spiral galaxy, resulting in an S0-like fast-rotating disk. This is followed by a phase of mini mergers that grows the outer spheroid stellar mass. UCDs and additional GCs are also accreted in this phase.  These mini mergers were not strong enough to destroy the central disk-like kinematics of the other tracers.}
    \label{fig:NGC3115_formationpath}
\end{figure*}

Based on our kinematic results from the stars and discrete tracers out to $\sim6.5\, R_\mathrm{e}$ (Sec. \ref{sec:results}) and on the evidence of weak spiral-arm structures in NGC 3115 \citep{Norris2006,Guerou2016}, we propose the following formation path for NGC 3115, which we schematically summarise in Fig. \ref{fig:NGC3115_formationpath}.

We propose that NGC 3115 was initially a spiral galaxy, with a star forming disk, that has undergone an early gas-rich minor merger of mass-ratio around 1:4-1:10. This merger event then shapes the current embedded fast-rotating disk 
that we observe from the kinematic profiles of the stars, red GCs and PNe.
A merger with this mass-ratio should not be strong enough to destroy the original disk structure but may be sufficient to cause a significant burst of star formation. 

We note that it is unlikely that NGC 3115 has undergone multiple minor mergers (more than two/three) of the mass-ratio 1:4-1:10 as that would result in an elliptical remnant without a prominent disk feature, i.e. not an S0 remnant \citep{Bournaud2005}. 

The stellar populations of the bulge and disk components of NGC 3115 indicate an ancient $12\, \mathrm{Gyr}$ old bulge. However, the disk is dominated by stars that formed some $9\, \mathrm{Gyr}$ ago, with a small contribution of younger stars
(see e.g. \citealt{Norris2006,Guerou2016}). 
\citet{Poci2019} also found evidence of a dominant old, metal-rich and kinematically cold stellar population in the disk component of NGC 3115, through their Schwarzschild orbit-superposition chemo-dynamical modelling technique, with signs of residual star formation that likely continued in the disk until $\sim4\, \mathrm{Gyr}$ ago when the galaxy exhausted its gas reservoir. Today NGC 3115 is a gas-poor system with very little cold molecular and atomic gas, and very little ongoing star-formation \citep{Li2011,Li2013}. The presence of this disk component rules out the occurrence of a major merger in recent times, as well as of multiple minor mergers of similar mass ratio to the early one as discussed above, as they would likely alter or destroy the old disk structure of NGC 3115. A passive evolution history of NGC 3115 is, therefore, favoured after the formation of its kinematically cold disk component. 
From the evidence from the stellar population results \citep{Guerou2016,Poci2019}, we constrain the epoch of the gas-rich minor merger to be around $9\, \mathrm{Gyr}$ ago.
We note that \citet{Usher2019} 
found several blue and red GCs of NGC 3115 to have ages around 9\, $\mathrm{Gyr}$, i.e. consistent with the estimated age of the merger. Thus some GCs may have formed during this gas-rich merger event.

At later epochs, NGC 3115 is likely influenced by mini mergers (mass-ratio $<$ 1:10) of dwarf galaxies which have been accreted onto the outer regions of the galaxy. These mini mergers would deposit GCs as well as stars in the outer regions giving rise to the
spheroid-like kinematics of the 
red GCs and PNe beyond $\sim2.5\, R_\mathrm{e}$, and perhaps of the 
blue GCs beyond $\sim5\, R_\mathrm{e}$.
According to \citet{Choksi2019} accretion is predicted to account for around half of the blue GC sub-population for a galaxy of stellar mass similar to NGC 3115. The old and metal-poor stellar population with hotter kinematics that characterizes the stellar halo of NGC 3115 as found by \citet{Poci2019} reinforces the proposed scenario that NGC 3115 late formation history is dominated by the accretion of low-mass satellite galaxies onto the outer regions of the galaxy. Specifically, \citet{Poci2019} suggested dry minor mergers to explain the old ages and metal-poor nature of the halo stars in NGC 3115, which they found to contribute to $\sim70\%$ of the stellar mass of NGC 3115. This significant {\it ex-situ} mass fraction in NGC 3115 is consistent with the proposed scenario that around half of the blue GCs, and some halo stars, may have likely been accreted from mini mergers with dwarf galaxies, which have contributed to the mass growth of the outer galaxy regions without altering the central cold disk kinematics, but causing the outer spheroid-like kinematics in NGC 3115.

We note that the above formation scenario is similar to that proposed by \citet{Arnold2011}. Based on GCs and stars from the SLUGGS survey, they described NGC 3115's inner regions having formed in an early gas-rich
dissipative phase. This is then followed by 1:15 to 1:20 mergers, which they refer to as minor mergers (here we use the term mini merger and reserve minor mergers for 1:4 to 1:10 mass ratios). 

\subsection{Origin of the UCDs in NGC 3115}
\label{sec:UCD_origin}
In Fig. \ref{fig:AllGCsplusUCDs_1D} and \ref{fig:AllGCsplusUCDs_1D_Original} of Sec. \ref{sec:GCvsUCDs_2D}, we have compared the velocities of the $12$ UCDs with the velocity field of the GCs. An inconsistency between the underlying velocity field of the GCs with that of the UCDs would suggest an accreted origin for the UCDs and their probable origin as nuclei of a tidally stripped dwarf. In fact, the blue GCs, that are typically associated with the accreted component, are rotating with NGC 3115, similarly to the stars, red GCs and PNe. Then an odd rotation of the UCDs would mean that they have been accreted at later epochs and the dynamical friction within the galaxy has still not brought them to the overall rotation of NGC 3115.

A consistency with the underlying velocity field of the GCs would, instead, favour an \textit{in-situ} and a massive GC origin of the UCDs, However, an accreted origin cannot be completely ruled out in this case as the UCD may have had the time to reach the dynamical equilibrium of the galaxy.

Our results show that five UCDs, namely UCD 1, 8, 15, 17a and 23, display an odd rotation with respect to the underlying GC velocity field, suggesting that they may have likely been accreted onto the galaxy. Specifically UCD 23 displays a counter-rotation with respect to the underlying GC velocity field.
Therefore, the odd rotation of these UCDs would point towards a possibly stripped dwarf origin. 
For all the remaining UCDs we find, instead, a good consistency with the underlying GC velocity field, suggesting an \textit{in-situ} origin.

In Sec. \ref{sec:NGC3115_formation}, we have discussed that the formation path of NGC 3115 is likely characterized by mini mergers (with mass-ratios $<$ 1:10) onto the outer regions of the galaxy at later epochs. Thus, the \textit{ex-situ} origin of our $5$ oddly rotating UCDs is a plausible scenario. 
However, we cannot reliably constrain the nature of these $5$ UCDs from only their inconsistent velocity with respect to the underlying GC velocity field. In fact, these UCDs could, instead, be massive GCs that have still been accreted at later epochs onto the outer regions of NGC 3115 \citep{Norris2014}. 
If we consider the inconsistencies that we have encountered with J14 (see Sec. \ref{sec:issues}), specifically when remeasuring the sizes of all the UCDs (see Sec. \ref{sec:ucd_sizes}), then the smaller sizes that we estimate for some of the UCDs (specifically for the three outliers) would make them more consistent with being massive GCs

For UCD 1, we have measured an old, $(12 \pm 2)\, \mathrm{Gyr}$, and metal-poor, $(-1.0 \pm 0.3)\, \mathrm{dex}$, stellar population (see Sec.~\ref{sec:stellar_population}).
Its velocity is higher than that of the underlying GC velocity field but its size, $\sim8\, \mathrm{pc}$, does not allow us to reliably determine whether it is a massive GC or a nucleus of a tidally stripped dwarf galaxy. 

The other seven UCDs have velocities which are consistent with the underlying GC velocity field and, for this reason, it seems more likely that these objects are indeed massive GCs \citep{Norris2014} which may have formed, for instance, during the gas-rich minor merger that characterized the assembly history of NGC 3115.

\section{Conclusions}
\label{sec:Conclusion}

In this paper, we have studied the kinematic properties of the nearest field S0 galaxy, NGC 3115, out to large galactocentric radii, i.e. $\sim6.5\, R_\mathrm{e}$. In order to reach such large distances, we have combined previous stellar kinematic studies (i.e. MUSE out to $\sim3\, R_\mathrm{e}$;  \citealt{Guerou2016} and SLUGGS out to $\sim3.5\, R_\mathrm{e}$;  \citealt{Arnold2011,Arnold2014}), with the large GC \citep{Pota2013,Forbes2017} and, recently published, PNe \citep{Pulsoni2018Survey} datasets. 
We have also performed new observations of a sample of compact objects around NGC 3115, which were initially identified by \citet{Jennings2014} as UCD candidates from {\it HST} imaging. We were able to spectroscopically confirm $7$ of these objects to be members of the galaxy from their radial velocity. These, combined with 5 UCDs from previous literature work, gives us a total of $12$ UCDs around NGC 3115 with radial velocities.

From the analysis of the {\it folded} 2D velocity and velocity dispersion maps and the corresponding 1D profiles,  
we find evidence of:

\begin{itemize}
    \item strong rotation, that is well aligned with the global photometric position angle of the galaxy, for each of the discrete tracers and stars (with a maximum misalignment of $\sim10\degr$ for the red GCs and PNe). This suggests an axi-symmetric system in dynamical equilibrium;
    \item the red GCs to have a maximum rotation amplitude between $\sim1-2\, R_\mathrm{e}$, similar to the PNe, while the blue GCs have their maximum rotation between $\sim3-5\, R_\mathrm{e}$. Beyond $\sim5\, R_\mathrm{e}$, the blue GC sub-population velocity profile possibly declines further, similarly to the red GCs and PNe beyond $2-2.5\, R_\mathrm{e}$;
    \item smooth kinematic transitions at $\sim0.2\, R_\mathrm{e}$ and $\sim2-2.5\, R_\mathrm{e}$ 
    from a dispersion-dominated bulge component ($V_\mathrm{rot}/\sigma < 1$) to a rotationally-supported disk component ($V_\mathrm{rot}/\sigma > 1$) and then to spheroid dominated kinematics ($V_\mathrm{rot}/\sigma < 1$) beyond $\sim2.5\, R_\mathrm{e}$. 
\end{itemize}

From the presence of spiral-arm remnant features in NGC 3115 \citep{Norris2006,Guerou2016} and from the comparison with simulations of galaxy formation, we suggest that NGC 3115 was originally a spiral galaxy that likely went through an early gas-rich minor merger of mass-ratio 1:4-1:10 some $9\, \mathrm{Gyr}$ ago. Such a merger modified (without destroying it) the original \textit{in-situ} disk structure of the spiral progenitor and shaped the current embedded kinematic disk of NGC 3115, forming perhaps a number of GCs. The kinematic profiles of the blue and red GC sub-populations
also support the case for such a minor merger.

The possible declining $V_\mathrm{rot}/\sigma$ profile beyond $\sim5\, R_\mathrm{e}$ of the blue GCs, as well as the declining profile of the red GCs and PNe beyond $\sim2-2.5\, R_\mathrm{e}$, suggests a late accretion phase of mini mergers (with mass-ratios $<$ 1:10) giving rise to spheroid-like kinematics. 
These mini mergers did not manage to disrupt the disk of the galaxy but likely contributed additional GCs and stars to the outer regions of the galaxy.  

The comparison of the UCD radial velocities with the underlying GC velocity field shows that five UCDs are oddly rotating, where one is even counter-rotating. This seems to point towards an \textit{ex-situ} and, possibly, a tidally stripped dwarf origin of these UCDs.

Overall, our study suggests a complex formation history for the nearest field S0 galaxy, NGC 3115. 
It also shows the importance of kinematic and spatially resolved stellar population studies of galaxies out to large galactocentric radii, as well as the need for more exhaustive simulation studies in order to better understand how mergers with different initial conditions influence the resulting large radii kinematics.

% ------------------- Acknowledgements ------------------- %
\section*{Acknowledgements}
We thank the anonymous referee for their comments and suggestions that have helped to improve this paper.
DF, WC, KB and AD acknowledge support from the Australian Research Council under Discovery Project 170102344.
AFM has received financial support through the Postdoctoral Junior Leader Fellowship Programme from `La Caixa' Banking Foundation (LCF/BQ/LI18/11630007).
SB acknowledges support from the Australian Research Council under Discovery Project 180103740. 
AJR was supported by National Science Foundation grant AST-1616710
and as a Research Corporation for Science Advancement Cottrell Scholar.
Some data presented herein were obtained at the W. M. Keck Observatory, which is operated as a scientific partnership among the California Institute of Technology, the University of California and the National Aeronautics and Space Administration. The Observatory was made possible by the generous financial support of the W. M. Keck Foundation. The authors wish to recognize and acknowledge the very significant cultural role and reverence that the summit of Maunakea has always had within the indigenous Hawaiian community. We are most fortunate to have the opportunity to conduct observations from this mountain.

%%%%%%%%%%%%%%%%%%%%%%%%%%%%%%%%%%%%%%%%%%%%%%%%%%



\bibliographystyle{mnras}
\bibliography{biblio} % if your bibtex file is called example.bib

%\begin{thebibliography}{99}
%\end{thebibliography}

%%%%%%%%%%%%%%%%%%%%%%%%%%%%%%%%%%%%%%%%%%%%%%%%%%

%%%%%%%%%%%%%%%%% APPENDICES %%%%%%%%%%%%%%%%%%%%%

\appendix

\section{UCD candidates}
\label{sec:UCD_candidates_images}
\begin{figure*}

\begin{subfigure}
\centering
\includegraphics[width=0.15\textwidth]{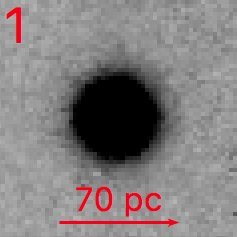}
\includegraphics[width=0.15\textwidth]{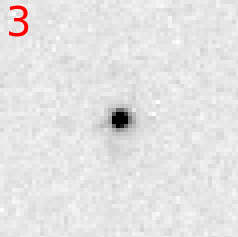}
\includegraphics[width=0.15\textwidth]{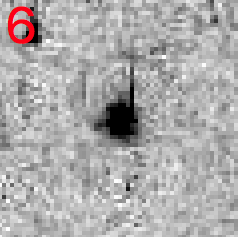}
\includegraphics[width=0.15\textwidth]{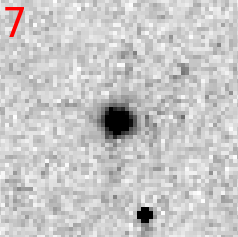}
\includegraphics[width=0.15\textwidth]{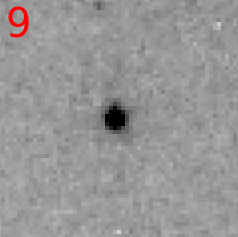}
\includegraphics[width=0.15\textwidth]{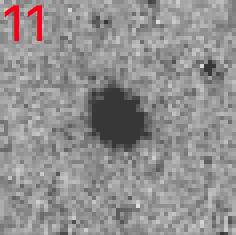}
\includegraphics[width=0.15\textwidth]{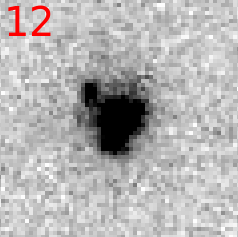}
\includegraphics[width=0.15\textwidth]{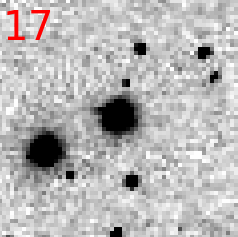}
\includegraphics[width=0.15\textwidth]{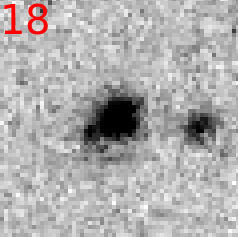}
\includegraphics[width=0.15\textwidth]{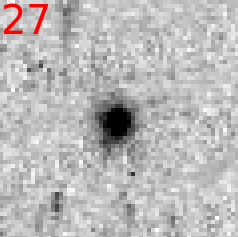}
\includegraphics[width=0.15\textwidth]{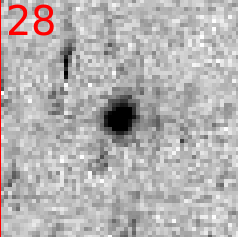}
\includegraphics[width=0.15\textwidth]{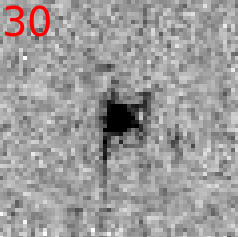}
\includegraphics[width=0.15\textwidth]{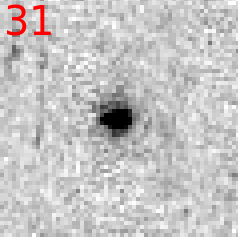}
\includegraphics[width=0.15\textwidth]{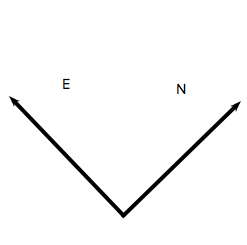}
\caption{Images of the $13$ UCD candidates from {\it HST}/ACS $g$-band images in our selected sample. The companion object in the field of UCD 17 has not been identified as either a GC or UCD of NGC 3115 by previous surveys. We refer to this ``bonus" object as UCD 17a. Each image has a size of $3\arcsec \times 3\arcsec$.
The numbers on the top-left side of each image represent the UCD ID as in J14. The arrow vectors show the North-East orientation of the {\it HST}/ACS images. The arrow vector in the UCD 1 image indicates a size of $1.5\arcsec$, corresponding to a physical size of $70\, \mathrm{pc}$ at the NGC 3115 distance (see Table \ref{tab:NGC3115_parameters}).}
\label{fig:GoodUCDs}
\end{subfigure}%

\medskip
\begin{subfigure}
\centering
\includegraphics[width=0.15\textwidth]{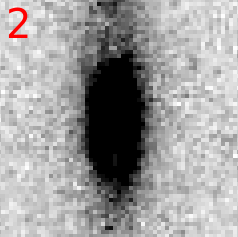}
\includegraphics[width=0.15\textwidth]{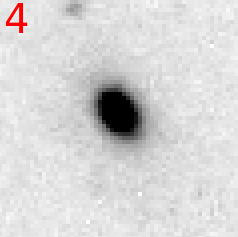}
\includegraphics[width=0.15\textwidth]{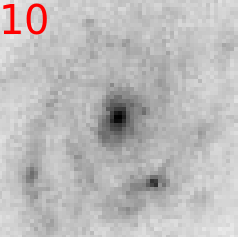}
\includegraphics[width=0.15\textwidth]{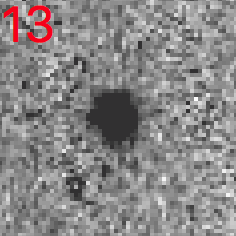}
\includegraphics[width=0.15\textwidth]{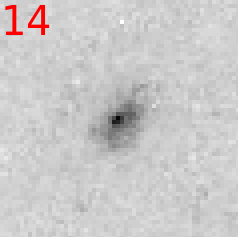}
\includegraphics[width=0.15\textwidth]{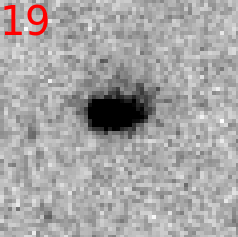}
\includegraphics[width=0.15\textwidth]{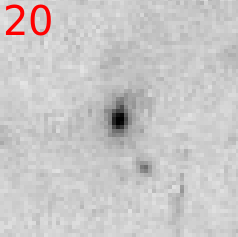}
\includegraphics[width=0.15\textwidth]{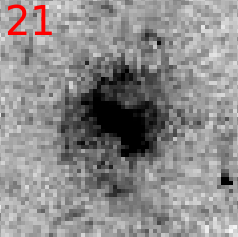}
\includegraphics[width=0.15\textwidth]{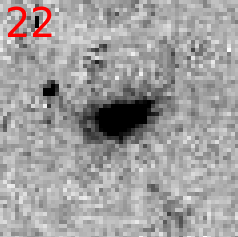}
\includegraphics[width=0.15\textwidth]{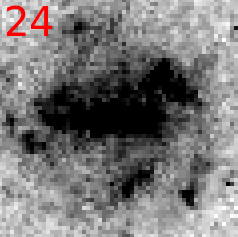}
\includegraphics[width=0.15\textwidth]{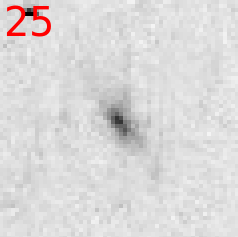}
\includegraphics[width=0.15\textwidth]{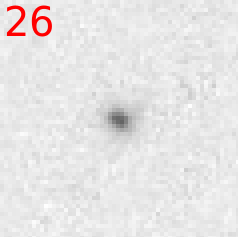}
\includegraphics[width=0.15\textwidth]{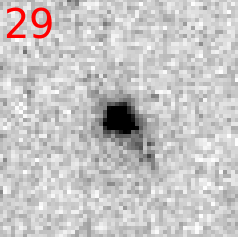}
\caption{Description as in Fig. \ref{fig:GoodUCDs}. 
Images of the $13$ UCD candidates from {\it HST}/ACS $g$-band images that we reject based on our visual selection criteria described in Sec. \ref{sec:sample_selection}. Specifically, \citet{Cantiello2015} also classified UCD 10 and UCD 20 as a background spiral galaxy and an interacting system with tidal features.}
\label{fig:BadUCDs}
\end{subfigure}

\medskip
\begin{subfigure}
\centering
\includegraphics[width=0.15\textwidth]{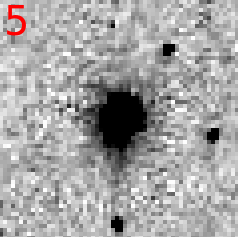}
\includegraphics[width=0.15\textwidth]{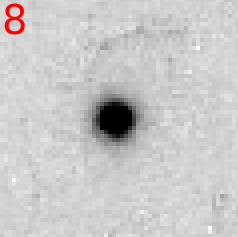}
\includegraphics[width=0.15\textwidth]{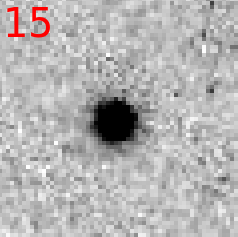}
\includegraphics[width=0.15\textwidth]{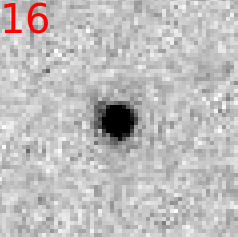}
\includegraphics[width=0.15\textwidth]{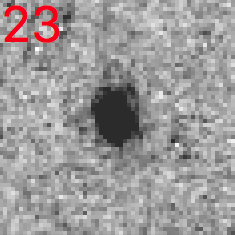}
\caption{Description as in Fig. \ref{fig:GoodUCDs}. {\it HST}/ACS $g$-band images of the $5$ objects which are spectroscopically confirmed members of NGC 3115 from their radial velocity measurements \citep{Pota2013} and later classified as UCDs by J14 from their size estimates, i.e. $R_{h} > 8\, \mathrm{pc}$.}
\label{fig:GoodUCDsVel}
\end{subfigure}
\end{figure*}

Figures \ref{fig:GoodUCDs}-\ref{fig:GoodUCDsVel} show the snapshots of the UCD candidates from J14, which we selected and rejected based on the visual inspection of the {\it HST}/ACS images described in Sec. \ref{sec:sample_selection}.

\section{Background objects}
\label{sec:background_objects}
From the spectral analysis in Sec. \ref{sec:kinematic_analysis}, we identify two of the observed UCDs, namely 11 and 12, as background objects. 
For both these objects, we do not identify any clear absorption or emission line feature in the sky-subtracted spectrum which is mostly noise. We only notice the presence of a very strong emission line at $\sim4940\, \angstrom$ for UCD 11 and at $\sim5154\, \angstrom$ for UCD 12. We do not recognize any of these emission lines as sky lines as they are also still the dominant ones in the object spectrum after the sky subtraction. 
For this reason, we assume that they possibly are the $H_{\beta}$ emission line coming from a more distant galaxy than NGC 3115. 
From the shift of this single emission line from the $H_{\beta}$ rest-wavelength, we estimate an heliocentric radial velocity $V = (4856.0 \pm 36.0)\, \mathrm{km\,s^{-1}}$ for UCD 11 and $V = (17550.0 \pm 15.0)\, \mathrm{km\,s^{-1}}$ for UCD 12, corresponding to a redshift of $z\sim0.016$ and $z\sim0.06$, respectively.

\begin{figure*}
\includegraphics[width=1.\textwidth]{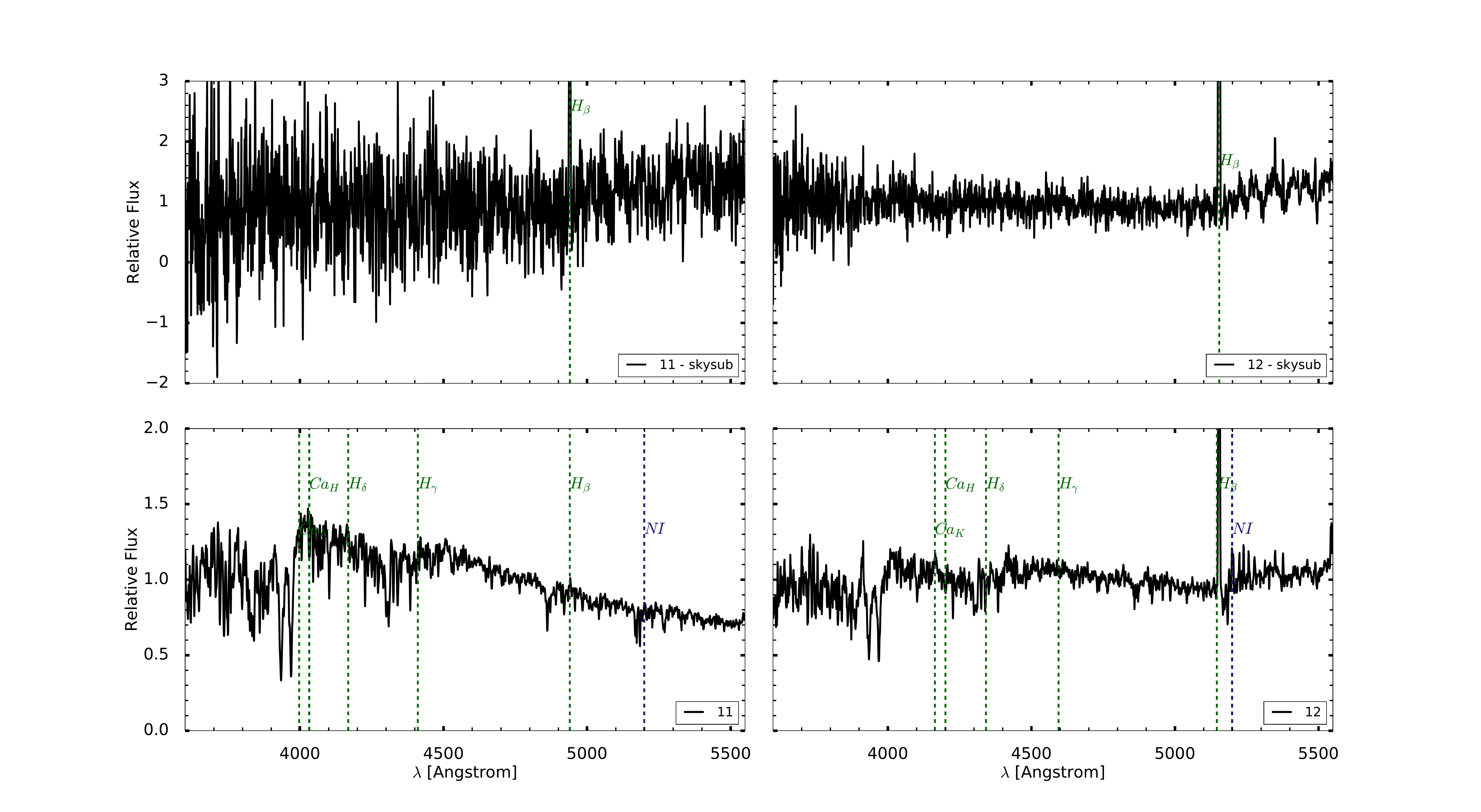}
\caption{From top to bottom: the sky-subtracted and non-sky subtracted spectra of UCD 11 and UCD 12, respectively. For UCD 11 and UCD 12, we label the strong visible emission line at $\sim4940\, \angstrom$ and at $\sim5154\, \angstrom$, which we assume to be red-shifted $H_{\beta}$. From the shift of this single emission line from the $H_{\beta}$ rest-wavelength, we measure an heliocentric radial velocity $V = (4856.0 \pm 36.0)\, \mathrm{km\,s^{-1}}$ and $V = (17550.0 \pm 15.0)\, \mathrm{km\,s^{-1}}$, which would locate the objects at $z\sim0.016$ and at $z\sim0.06$, respectively. 
As dark blue dashed line we label the sky emission line $\mathrm{NI}=5199\, \angstrom$. The other labelled main absorption lines are shifted to the expected wavelength of the objects if they were receding with the assumed heliocentric velocity.}
\label{fig:background_objects}
\end{figure*}

In Fig. \ref{fig:background_objects} we show the sky-subtracted (top panels) and non sky-subtracted (bottom panels) spectra of UCD 11 (left) and UCD 12 (right). In the sky-subtracted spectra, we observe the previously mentioned prominent emission lines of UCD 11 and 12 and the noise dominating in all other regions of the spectra.
In the non sky-subtracted spectra, we observe the main absorption features that we also label in Sec. \ref{sec:kinematic_analysis}, but here they lie at their respective rest-wavelengths, meaning that they are $z\sim0$ features not coming from the distant objects.
We highlight here that the observations (see Sec. \ref{sec:Observations}) were conducted during a bright night, therefore these absorption lines could be a result of the scattered light from the Moon.

\section{Comments on Jennings et al. (2014)}
\label{sec:issues}
Here, we briefly mention some of the issues/inconsistencies with the UCD candidates described in J14 that we have encountered during our analysis. 
\begin{itemize}
    \item We find that there is a small offset in the RA and DEC coordinates of the UCDs given in J14 from the centre of each UCD in both $g$ and $z$ filters of the \textit{HST}/ACS images. Specifically, we find that roughly half of the UCDs have RA and DEC coordinates offset by $\sim0.5\arcsec$ in either the $g$ or $z$ filter.
    In Table \ref{tab:UCD_Candidates_Table}, we list our newly derived RA and DEC coordinates from the {\it HST}/ACS $g$-band images for all $31$ UCD candidates listed in J14.
    \item The list of $31$ UCD candidates given in J14 contained background galaxies which can be clearly identified from the visual inspection of the {\it HST}/ACS images (see Fig. \ref{fig:BadUCDs}). From these images, we selected our sample of UCD candidates, i.e. $19$ out of the initial $31$ (see Fig. \ref{fig:GoodUCDs} and \ref{fig:GoodUCDsVel}). Additionally, from the spectroscopic analysis (see Sec. \ref{sec:kinematic_analysis}), we identify two of the UCDs as likely background objects located at higher redshifts than NGC 3115 (see Appendix. \ref{sec:background_objects}). It cannot be excluded that other UCD candidates may also be background objects.
    \item In Sec. \ref{sec:ucd_sizes}, we remeasure the sizes (i.e. half-light radius, $R_{h}$) of all $31$ UCD candidates by fitting a S\'ersic function to each object. For three objects, which belong to our selected sample of UCD candidates (see Fig. \ref{fig:GoodUCDs}), we measure much smaller sizes than J14 (see Table \ref{tab:UCDs_Size_Table}). From the residual images (see Fig. \ref{fig:UCD367_QFitsViewFit}) and compactness of the objects, it seems unlikely that we are underestimating their true sizes by such a large factor.
    \end{itemize}
    
\begin{table}
\caption{UCD sizes of the three objects (UCDs 3, 6, 7), included in our selected sample of Fig. \ref{fig:GoodUCDs}, for which we find the largest difference with the J14 estimates. \textit{Column 1:} UCD ID, \textit{Column 2:} UCD $R_{h}$ measured from fitting a S\'ersic function to each object, \textit{Column 3:} UCD $R_{h}$ from J14. Each object is assumed to lie at the NGC 3115 distance.} 
\centering
\begin{tabular}{|c|c|c|}\hline\hline
UCD & $R_{h}\, \mathrm{(pc)}$ [This work] & $R_{h}\, \mathrm{(pc)}$ [J14] \\ \hline
3   & 5 & 70.6 \\
6   & 5 & 70.6 \\
7   & 6 & 73.6 \\
\hline \hline
\end{tabular}
\label{tab:UCDs_Size_Table}
\end{table}

%%%%%%%%%%%%%%%%%%%%%%%%%%%%%%%%%%%%%%%%%%%%%%%%%%

% Don't change these lines
\bsp	% typesetting comment
\label{lastpage}
\end{document}